\documentclass[sigconf,nonacm]{acmart}

\AtBeginDocument{%
  }

\setcopyright{acmlicensed}
\copyrightyear{2025}
\acmYear{2025}
\acmConference[CUG 2025 ]{2025 Cray User Group Meeting}{May 04--10, 2025}{Jersey City, NJ}

\begin{document}

\title{Scaling MPI Applications on Aurora}

\author{Huda Ibeid}
\email{huda.ibeid@intel.com}
\affiliation{%
\institution{Intel Corporation}
\city{Santa Clara}
\state{California}
\country{USA}}

\author{Anthony-Trung Nguyen}
\email{anthony.d.nguyen@intel.com}
\affiliation{%
\institution{Intel Corporation}
\city{Santa Clara}
\state{California}
\country{USA}}

\author{Aditya Nishtala}
\email{aditya.nishtala@intel.com}
\affiliation{%
\institution{Intel Corporation}
\city{Santa Clara}
\state{California}
\country{USA}}

\author{Premanand Sakarda}
\email{premanand.sakarda@intel.com}
\affiliation{%
\institution{Intel Corporation}
\city{Santa Clara}
\state{California}
\country{USA}}

\author{Larry Kaplan}
\email{lkaplan@hpe.com}
\affiliation{%
\institution{Hewlett Packard Enterprise}
\city{Shoreline}
\state{WA}
\country{USA}}

\author{Nilakantan Mahadevan}
\email{nilakantan.mahadevan@hpe.com}
\affiliation{%
\institution{Hewlett Packard Enterprise}
\city{Springs}
\state{Texas}
\country{USA}}

\author{Michael Woodacre}
\email{woodacre@hpe.com}
\affiliation{%
\institution{Hewlett Packard Enterprise}
\city{Springs}
\state{Texas}
\country{USA}}

\author{Victor Anisimov}
\email{vanisimov@anl.gov}
\affiliation{%
\institution{Argonne National Laboratory}
\city{Lemont}
\state{Illinois}
\country{USA}}

\author{Kalyan Kumaran}
\email{kumaran@anl.gov}
\affiliation{%
\institution{Argonne National Laboratory}
\city{Lemont}
\state{Illinois}
\country{USA}}

\author{JaeHyuk Kwack}
\email{jkwack@anl.gov}
\affiliation{%
\institution{Argonne National Laboratory}
\city{Lemont}
\state{Illinois}
\country{USA}}

\author{Vitali Morozov}
\email{morozov@anl.gov}
\affiliation{%
\institution{Argonne National Laboratory}
\city{Lemont}
\state{Illinois}
\country{USA}}

\author{Servesh Muralidharan}
\email{servesh@anl.gov}
\affiliation{%
\institution{Argonne National Laboratory}
\city{Lemont}
\state{Illinois}
\country{USA}}

\author{Scott Parker}
\email{sparker@anl.gov}
\affiliation{%
\institution{Argonne National Laboratory}
\city{Lemont}
\state{Illinois}
\country{USA}}

\renewcommand{\shortauthors}{}

\begin{abstract}
  The Aurora supercomputer, which was deployed at Argonne National Laboratory in 2024, is currently one of three Exascale machines in the world on the Top500 list. The Aurora system is composed of over ten thousand nodes each of which contains six Intel Data Center Max Series GPUs, Intel's first data center-focused discrete GPU, and two Intel Xeon Max Series CPUs, Intel's first Xeon processor to contain HBM memory. To achieve Exascale performance the system utilizes the HPE Slingshot high-performance fabric interconnect to connect the nodes. Aurora is currently the largest deployment of the Slingshot fabric to date with nearly 85,000 Cassini NICs and 5,600 Rosetta switches connected in a dragonfly topology. The combination of the Intel powered nodes and the Slingshot network enabled Aurora to become the second fastest system on the Top500 list in June of 2024 and the fastest system on the HPL MxP benchmark. The system is one of the most powerful systems in the world dedicated to AI and HPC simulations for open science. This paper presents details of the Aurora system design with a particular focus on the network fabric and the approach taken to validating it. The performance of the systems is demonstrated through the presentation of the results of MPI benchmarks as well as performance benchmarks including HPL, HPL-MxP, Graph500, and HPCG run on a large fraction of the system. Additionally results are presented for a diverse set of applications including HACC, AMR-Wind, LAMMPS, and FMM demonstrating that Aurora provides the throughput, latency, and bandwidth across system needed to allow applications to perform and scale to large node counts and providing new levels of capability and enabling breakthrough science.  
  
\end{abstract}



\maketitle

\section{Introduction}\label{Introduction}

The Aurora system was designed by Intel and HPE and deployed in 2024 by the Argonne Leadership Computing Facility at Argonne National Lab for the purpose of providing large scale computational resources for a new scale of open scientific computing for breakthrough science. The system is designed to support a wide variety of different simulation and AI applications developed and run members of scientific community who have been grants of time on the system. As an exascale-class system Aurora is one of the most powerful computing systems ever built, representing major step forward in computational capabilities for the scientific community.

As HPC systems have grown larger and more powerful the complexities in deploying and utilizing these systems has increased. Aurora is built from over 60,000 GPUs, 20,000 CPUs, 80,000 NICs, 5,600 switches, and 200,000 network links along with numerous other components that must all be functioning correctly to in order for the system to be fully functional. In addition, Aurora is the largest deployment to date of the HPE Slingshot fabric and the first large scale deployment of Intel's first discrete data center GPU, each of which contains over 100 billion transistors. In order to utilize the Aurora system effectively applications need to be able to not only effectively utilize the systems CPUs and GPUs at the node level but applications targeting the system are expected to run on thousands of nodes with many requiring close coordination across the nodes using the Slingshot fabric in order to solve their scientific problems. 

This paper will present an overview of the Aurora system (Section 2) with a particular focus on the Slingshot network (Section 3), and a discussion of the approach utilized to validate the functionality and performance of the Slingshot network (Section 4). Followed by results from MPI benchmarks demonstrating the networks performance along with results from several  HPC benchmarks (HPL, HPL-MxP, Graph500, HPCG) run at large scale, and finally scaling results for several real world applications (HACC, Nekbone, AMR-Wind, LAMMPS, FMM) demonstrating the capability of running scientific applications at large scale on the machine (Section 5).
Conclusions are then provided (Section 6).
\section{System Design}\label{System}

\begin{figure}[htbp]
    \centering
     \includegraphics[width=0.85\linewidth]{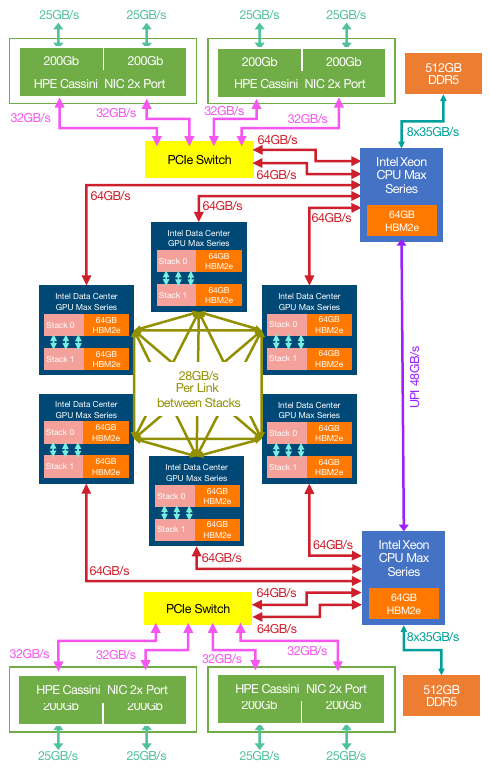}
     \caption{Aurora Node}
     \label{fig:Aurora_Node}
     \Description{Aurora Node} 
 \end{figure}

All of the nodes on the Aurora system are identical and have two Intel Xeon Max Series CPUs with HBM, also known as Sapphire Rapids (SPR), and six (6) Intel Data Center Max GPUs, also known as Ponte Vecchio (PVC) as show in Figure ~\ref{fig:Aurora_Node}.

The Intel Xeon Sapphire Rapids CPUs on Aurora have 52 cores and contain of 64 GB HBM2e memory in addition to being connected to 512 GB of DDR5 memory. The CPUs on a node are in a dual socket configuration and are connected with Intel's UPI interconnect. In addition to Intel SPR CPUs, each Aurora node contains six Intel PVC GPUs. Aurora is the earliest and largest deployment of the Intel PVC GPU which was designed to meet the performance needs of HPC and AI applications. 

The basic building block of the PVC GPU is the Xe-Core, which consists of eight vector and matrix engines, with a combined register file size of 512 KB. The vector unit is 512 bits wide (8-wide for double precision floating point (FP64)) and supports fused multiply-add (FMA) operations. The matrix unit is designed to support mixed-precision matrix multiplications, commonly used in AI applications. The engines within the Xe-Core share a large 512 KB high-bandwidth L1 data cache which can be partially configured as highly banked Shared Local Memory, serving as a software-managed scratch pad. On PVC 64 Xe-Cores are grouped into an Xe-Stack which share the Last Level Cache (LLC), HBM2e memory controllers, high-speed Stack-to-Stack interconnect within the same silicon substrate, and the Xe-link high-speed coherent fabric for remote GPU-to-GPU communication. The LLC in each Xe-Stack is a 192 MiB memory-side cache connected to its dedicated HBM2e memory stacks. Two Xe-Stacks, along with their local HBM2e memory stacks, are integrated into a package having 128 Xe-Cores and 128 GB of HBM2e memory.  GPUs on a node are connected in all-to-all manner with dedicated Xe-Links, with each capable of 28 GB/s of theoretical bandwidth. The GPUs are directly connected to the CPUs via PCIe Gen5 x16 links offering 64 GB/s design bandwidth.

Each Aurora node contains eight Slingshot Cassini NICs, four of which are connected by PCIe Gen 4 links through a PCIe switch to each CPU. Though the NICS are not connected directly to the GPUs, data may pass from the PVC GPUs directly to the Cassini NICs without being loaded into CPU memory. The NICs connect to the Slingshot fabric which connects 10,624 Aurora nodes together to form the Aurora system. Further details of the Aurora Slingshot network are discussed in the following section. The Aurora storage system is composed of a novel 260 PB Distributed Asynchronous Object Store (DAOS) that uses high performance flash memory with a peak bandwidth of 31 TB/s along with a 91 TB Lustre file system capable of a peak bandwidth of 650 GB/s and a smaller 12 PB Lustre file system holding user home directories. The overall specifications for the Aurora system are shown in Table ~\ref{tab:aurora_compute_spec}.

\begin{table}[htbp]
\centering
\caption{Aurora Aggregate Specifications.}
\begin{tabular}{lc}
\toprule
Nodes & 10,624 \\
No. of CPUs & 21,248 \\
No. of GPUs & 63,744 \\
DDR5 Memory Capacity & 10.62~PB \\    
DDR5 Memory Bandwidth & 5.31~PB/s \\  
HBM2e Memory Capacity & 9.52~PB \\    
HBM2e Memory Bandwidth & 147.46~PB/s \\  
Injection Bandwidth & 2.12~PB/s \\    
Global Bandwidth & 1.37~PB/s \\ 
\bottomrule
\end{tabular}
\label{tab:aurora_compute_spec}
\end{table}

\section{Networking}\label{Networking}

Aurora is built using the HPE Cray EX supercomputer hardware with Slingshot-11~\cite{cug2022slingshot}. Slingshot-11 consists of Rosetta-1 (Switch) and Cassini-1 (NIC). The rest of this section gives a detailed description of the fabric. 

\subsection{Description of Aurora Fabric}
\begin{figure*}[htb]
    \centering
     \includegraphics[width=0.94\linewidth]{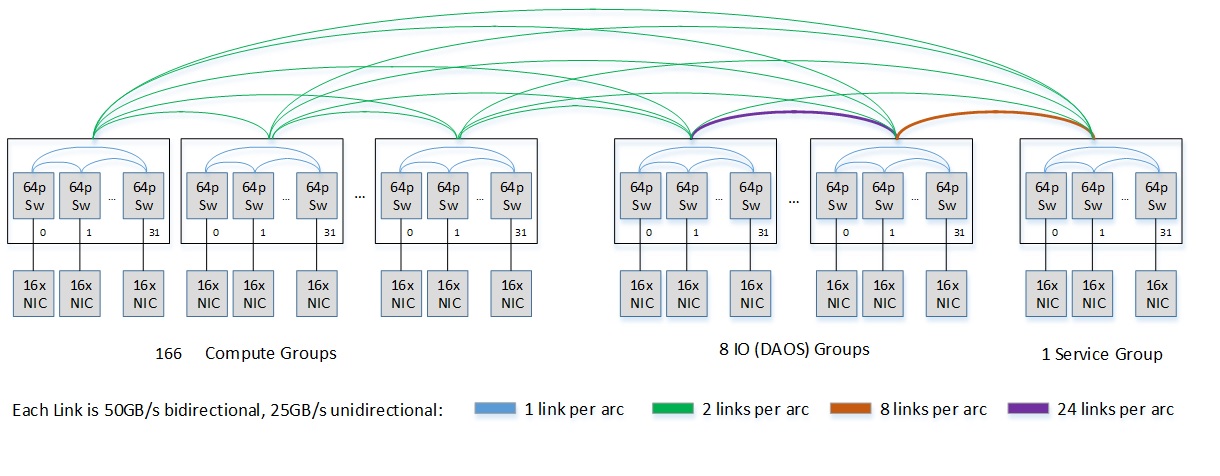}
     \caption{Aurora Fabric Topology.}
     \label{fig:Aurora_Fabric}
     \Description{Aurora Slingshot DragonFly group.}
 \end{figure*}

The Aurora network shown in Figure~\ref{fig:Aurora_Fabric} is composed of 166 compute groups, 8 storage groups, and 1 service group connected in a dragonfly topology~\cite{4556717}. Aurora’s network architecture follows a single dimension dragonfly topology that uses all-to-all local groups connected to each other through global links. Each compute group corresponds to a single HPE Cray EX cabinet and contains 32 switches that are all-to-all connected using one network link between each pair of switches. Each switch connected to two nodes, each having eight endpoints, for a total of 512 endpoints per compute group. The storage and service groups are each spread across multiple River cabinets and contain 32 switches that are all-to-all connected using one network link between each pair of switches. All intra-group cables in the compute cabinets are electrical and Intra-group cables within the storage groups consist of a mix of electrical and optical. As in the compute groups, each switch connects to 16 endpoints, for a maximum of 512 endpoints per group. The 166 compute groups contain 84,992 endpoints in the system that deliver 2.12 PB/s of injection bandwidth. Individual compute groups are connected all-to-all at the global level with every pair of compute groups connected by 2 links.  This provides a total of 1.38 PB/s of global bandwidth and 0.69 PB/s of global bisection bandwidth between compute groups.  Each compute group connects to each of the 9 non-compute groups (1 Service group and 8 storage/DAOS groups) with 2 links. Finally, the 8 I/O DAOS groups require significant bandwidth among themselves and, therefore, have 24 links between each pair of DAOS groups. The global network between groups uses optical cables, each of which carries a pair of network links for a bandwidth of 50 GB/s/dir. Bundles of one optical cable are used between groups. At most, three switch-to-switch hops are required to route a packet minimally to any destination: one hop in the source group, one global hop between groups, and one hop in the destination group.  Aurora has the largest existing deployment of a Slingshot-11 interconnect in a dragonfly topology with over 300,000 ports (241,428 fabric ports and  87404 edge ports).
 
The system provides fully dynamic routing. Adaptive routing decisions are made on a packet-by-packet basis for all unordered traffic. Where ordered delivery is required, dynamic adaptive routing decisions are made for each new destination, and such decisions persist for the lifetime of an active flow, i.e., a new decision on how to route traffic to a given destination will be made whenever no traffic is pending to that destination. The network uses fully adaptive routing for bulk data delivery. Ordered delivery is used when required by the transport layer protocol (e.g., for ordering MPI envelopes). In the absence of contention, all traffic is routed minimally, taking at most one hop in the source group, one global hop, and one hop in the destination group. In the presence of congestion, the adaptive routing mechanism can cause packets to route non-minimally to avoid the congestion. The system implements novel and highly effective congestion management which reduces interference between jobs. The switch hardware will detect congestion, identify its causes, and determine whether traffic flowing through a congested point is contributing to the congestion or is a victim of it. The switch hardware applies stiff back pressure to congesting traffic, limiting injections by members of an incast to their fair share of bandwidth. All traffic not contributing to the congestion is unaffected by the back-pressure and is free to pass any blocked packets.

The system supports QoS on all network traffic for performance isolation across different classes of traffic, independent of the congestion control. Network traffic is tagged with a traffic class according to the issuing queue. Traffic shaping in the NICs and the switches operate on these classes. In the absence of bandwidth contention, the arbiters select packets to forward based on their traffic class and the credits available to that class. Slingshot supports minimum and maximum bandwidths for each traffic class. If a class does not use its minimum bandwidth, other classes may use it, but no class will consume more than its maximum bandwidth. Slingshot also supports priority-based traffic scheduling, to prioritize traffic from a given class above that of other classes until the classes’ credits have been consumed. The duration over which the prioritization is averaged is a tunable parameter.

Slingshot supports a virtual channel mechanism that is specific to routing and hidden beneath traffic classes. Virtual channels are not visible to users. The network design provides significant flexibility to site administrators. The system is configured to guarantee bandwidth available for I/O traffic. Jobs run via a specific queue could be allocated to a traffic class with high minimum bandwidth guarantees, dedicating most of the network resources to such jobs when they are present. Low latency operations such as barriers and reductions could run in a high-priority traffic class. Debugging and performance tools could operate within a traffic class with a low maximum bandwidth.

\subsection{Rosetta Switch}
Rosetta is a 64-port Ethernet switch,developed by HPE, implemented as a large monolithic (685 mm2) ASIC fabricated in the TSMC 16 nm FinFet process and housed in a 62.5 mm package. All main switching logic utilizes an 850 MHz clock, which results in a typical power dissipation of 160 watts and a maximum of around 300 watts. 

\subsection{Cassini NIC}
Cassini is a 200 Gbps HPC NIC ASIC chip developed by HPE. The host interface is PCIe Gen4 with support for extended speed mode (where supported by the CPU or GPU). The network link port conforms to the 200 Gbps (4×50 Gbps PAM 4) and 100 Gbps (4×50 Gbps NRZ) Ethernet standards. The link between a Cassini NIC and a  Rosetta switch can carry standard Ethernet frames and Slingshot protocol frames simultaneously. On top of those base transports Rosetta can carry IPv4 and IPv6 packets, including RoCE, plus RDMA packets for MPI, one-sided operations, storage, and low latency collectives.

The Cassini drivers and libaries provide upper layers of software with a reliable RMA and messaging transport conforming to the OFI libfabric API~\cite{libfabric}.  When a packet is lost, or resources are exhausted, the transport will take the actions required to complete the operation. Cassini provides two reliability models: a simplified or “restricted” model for idempotent operations and an “unrestricted” model for operations that manipulate target state.  The restricted model is connection-less. The unrestricted model uses dynamically allocated connections that persist only while a specific operation is in progress.  It stores the result of an operation, checking its store for the result before executing a retry, and releasing the result as the operation completes.  

\subsection{Enhanced Link Functionality}

The Slingshot architecture provides optimizations within the network stack designed to improve network efficiency. These enhancements, driven by HPC and data analytics workloads,  increase the performance of the data link layer. These improvements benefit both HPC and Ethernet applications since they enable greater throughput, permit a higher transaction rate, and improve reliability. Enhancements include reduced inter-packet gap, optimized packet headers, credit-based flow control, link-level retry, and degraded link operation (i.e. continued operation with two or three of the four lanes disabled). Enhanced functionality is negotiated between pairs of devices using Link Layer Discovery Protocol (LLDP).

\subsection{Fabric Management}
The HPE Slingshot Fabric Manager shown in Figure \ref{fig:Slingshot_Fabric_Management}  includes a suite of software which configures, manages, and monitors the network. It runs on a fabric manager node, an external server,  and communicates with the switches over the out-of-band management network. The Fabric Manager node is an Intel Ice Lake server consisting of an Intel(R) Xeon(R) Gold 5320 CPU with 256GB RAM with 11TB of storage running SUSE Linux. For resiliency, there are two fabric manager nodes configured as an Active-Standby Cluster. Additional details on Fabric Manager software configuration aspects can be found in software section   \ref{Fabric_Management}.

\subsection{Algorithmic MAC Address}
To support efficient low latency forwarding of frames within the network, Slingshot uses algorithmically generated fabric addresses assigned to physical ports (HSN NICs) based on the network topology, enabling low latency and interval routing to be used within the High-Speed Fabric.

\subsection{Address Resolution Protocol (ARP)}
Aurora uses a Static/Permanent ARP method that avoids broadcast and multicast traffic. In addition, this method results in better job startup time.  In environments where the IP address to MAC mapping is constant, it is recommended to load the ARP entries of all compute nodes HSN IPs as static/permanent entries in all compute nodes. This will result in ARP entries in the host cache which are not invalidated. This is done as a part of the compute node boot sequence. The ARP cache on the host system is appropriately modified based on the total number of endpoints in the system.

\subsection{Fabric Validation}\label{fabric_validation}

\begin{figure*}[htb]
    \centering
     \includegraphics[width=0.94\linewidth]{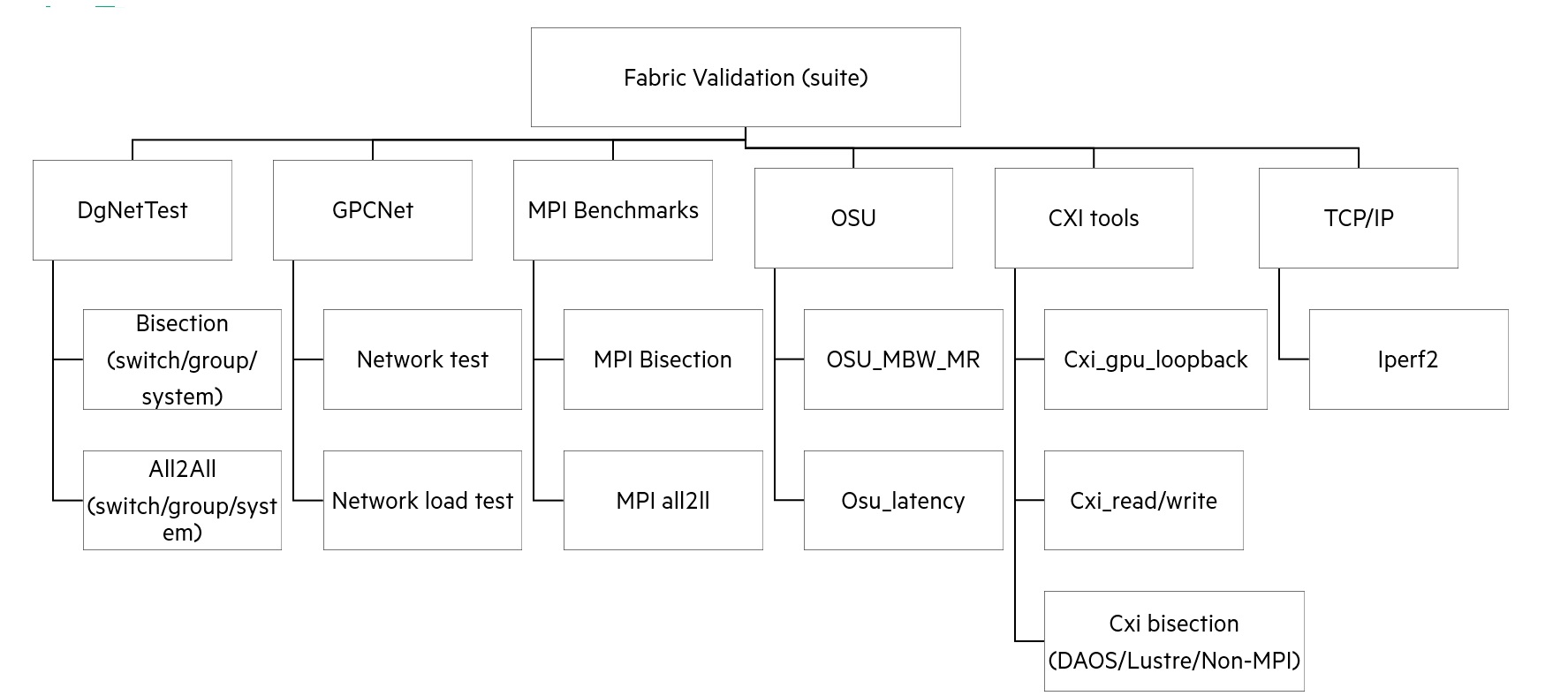}
     \caption{Aurora Fabric Validation.}
     \label{fig:Aurora_Fabric_Validation}
     \Description{Aurora Fabric Validation.}
 \end{figure*}
 
This section provides an overview of the various performance validation tools used to test the network on Aurora. These tools as shown in Figure~\ref{fig:Aurora_Fabric_Validation}
encompass tools developed by HPE and open source alternatives. By following a systematic process, failures and low performing nodes or groups can be
identified and isolated at an early stage. The underlying principle of this process is that the overall system health depends on the health of
all groups. To ensure a group’s health, it is essential that all switches and endpoints within that group are also healthy. 
\subsubsection{MPI all2all tests}\label{all2all}
MPI all2all  is considered as a vital pre-flight test prior to running large scale HPC and AI Benchmarks like HPL and HPL-MxP. It is important for the following reasons:
\begin{itemize}
    \item Diagnosing any connectivity issues
    \item Exposing any hardware issues that can impact the run and can be helpful to isolate issues on nodes
    \item Resolving performance bottlenecks 
    \item Tesing different transfer sizes with benchmarks and applications using varying transfer sizes
    \item Diagnosing fabric issues 
\end{itemize}

\begin{figure}[htb]
    \centering
     \includegraphics[width=0.9\linewidth]{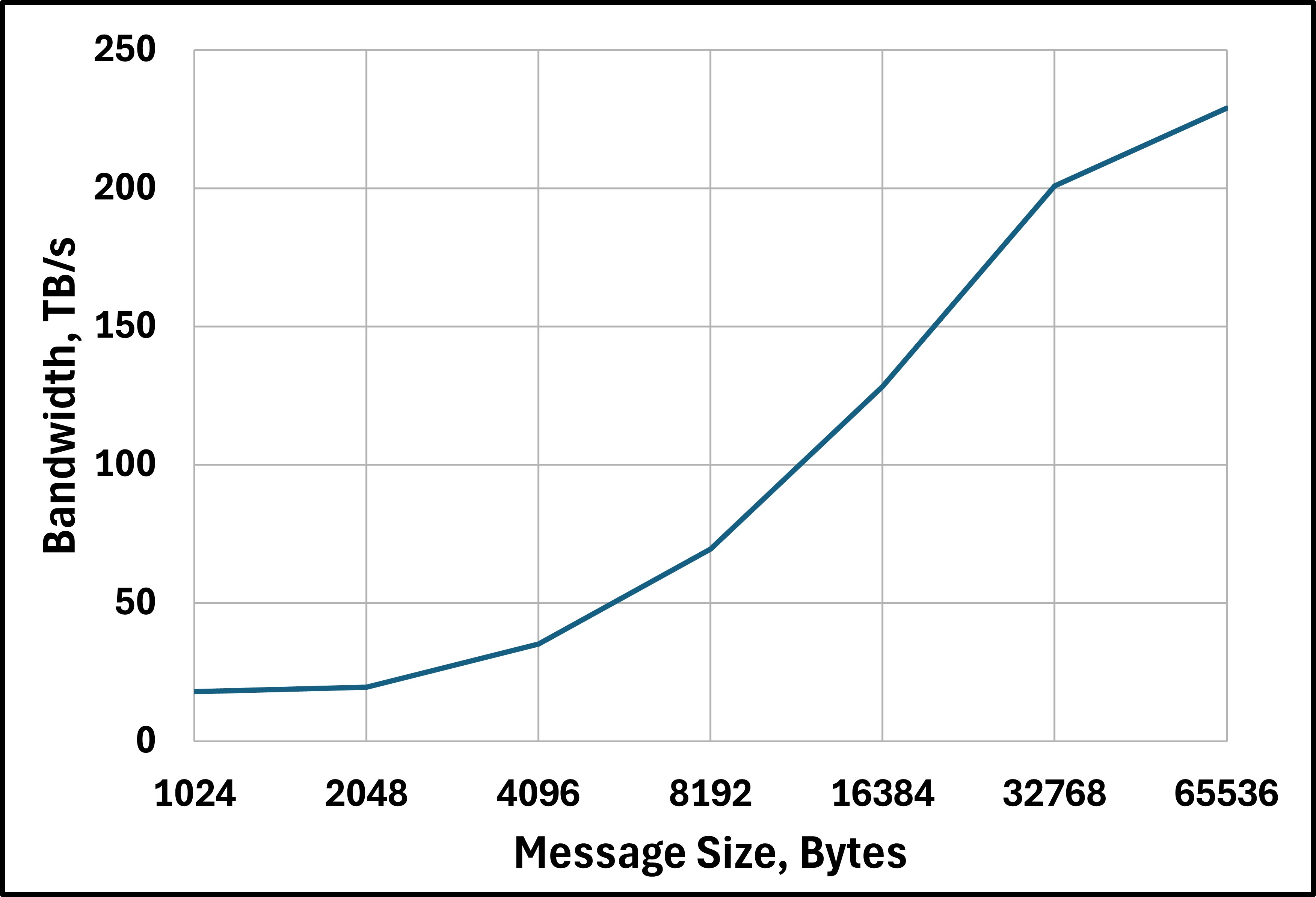}
     \caption{Fabric validation using an all2all communication benchmark across 9,658 nodes (77,264 NICs) with 16 processes per node (PPN=16)}
     \label{fig:all2all-scale}
     \Description{Graph with all2all with 9658 nodes and 77264 NICs.}
 \end{figure}

 Figure~\ref{fig:all2all-scale} shows an example MPI all2all test at a scale of 9658  nodes (90 percent of overall system) that was run as fabric validation prior to a successful HPL-MxP benchmark run. The smooth scaling for various transfer sizes achieves a peak aggregate bandwidth of 228.92 TB/s. The nodes that passed the tests were then subsequently used for HPL MxP benchmark explained in \ref{HPL-MxP} achieving the \#1 score in the world.

 \subsubsection{GPCnet tests}\label{GPCNET}
 GPCNet~\cite{GPCNet} is an MPI benchmark which includes the following capabilities:
 \begin{itemize}
     \item Natural Ring Communication Pattern - All processors communicate with their neighbors simultaneously based on adjacent MPI ranks.
     \item Random Ring Communication Pattern - Processors are paired with random nodes that do not live on a physical neighbor in the machine.
 \end{itemize}

\begin{figure}[htb]
    \centering
     \includegraphics[width=0.9\linewidth]{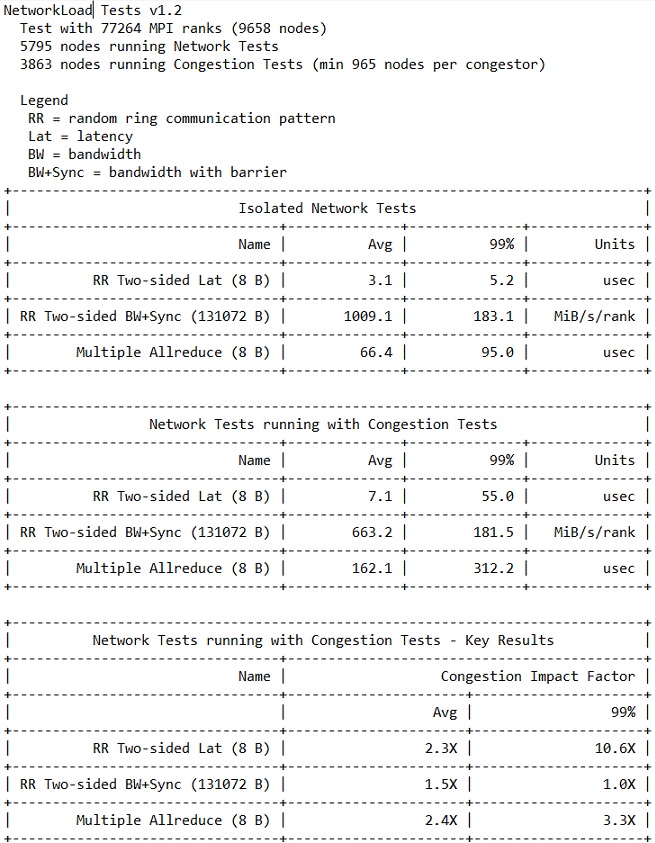}
     \caption{Fabric Validation using GPCNet network  test}
     \label{fig:GPCNET_Scale}
     \Description{Graph with GPCNET network  test with 9658 nodes and 77264 NICs.}
 \end{figure}

These tests are run in an isolated fashion by GPCNet and then they are run again concurrently with traffic patterns that deliberately try to generate congestion.  Figure~\ref{fig:GPCNET_Scale} shows the result of GPCNet (network load test) at a scale of 9658  nodes (90 percent of overall system). This is the largest scale at which GPCnet has been executed successfully.

Nodes that completed MPI all2all and GPCnet tests successfully were used for the HPL Benchmark as explained in \ref{HPL} resulting in Aurora achieving Exascale performance. 

\subsubsection{OSU MBW MR}\label{osu}
The OSU Benchmark suite~\cite{OSU} is a collection of MPI tests  that  collect key performance metrics (bandwidth and latency) for HPC fabrics on a variety of traffic patterns.
They can operate on any of the standard MPI implementations. OSU Micro-Benchmarks provide several different types of benchmarks that
can be used to test and measure HPC network performance:
\begin{itemize}
    \item Point-to-Point MPI Benchmarks: Latency, multi-threaded latency, multi-pair latency, multiple bandwidth / message rate test bandwidths,
bidirectional bandwidth.
\item One-sided MPI Benchmarks: one-sided put latency, one-sided put bandwidth, one-sided put bidirectional bandwidth, one-sided get
latency, one-sided get bandwidth, one-sided accumulate latency.

\end{itemize}

For fabric validation, osu\_mbw\_mr test is used for bandwidth and osu\_multi\_lat test is used for latency.
 
\begin{figure}[h]
    \centering
     \includegraphics[width=0.9\linewidth]{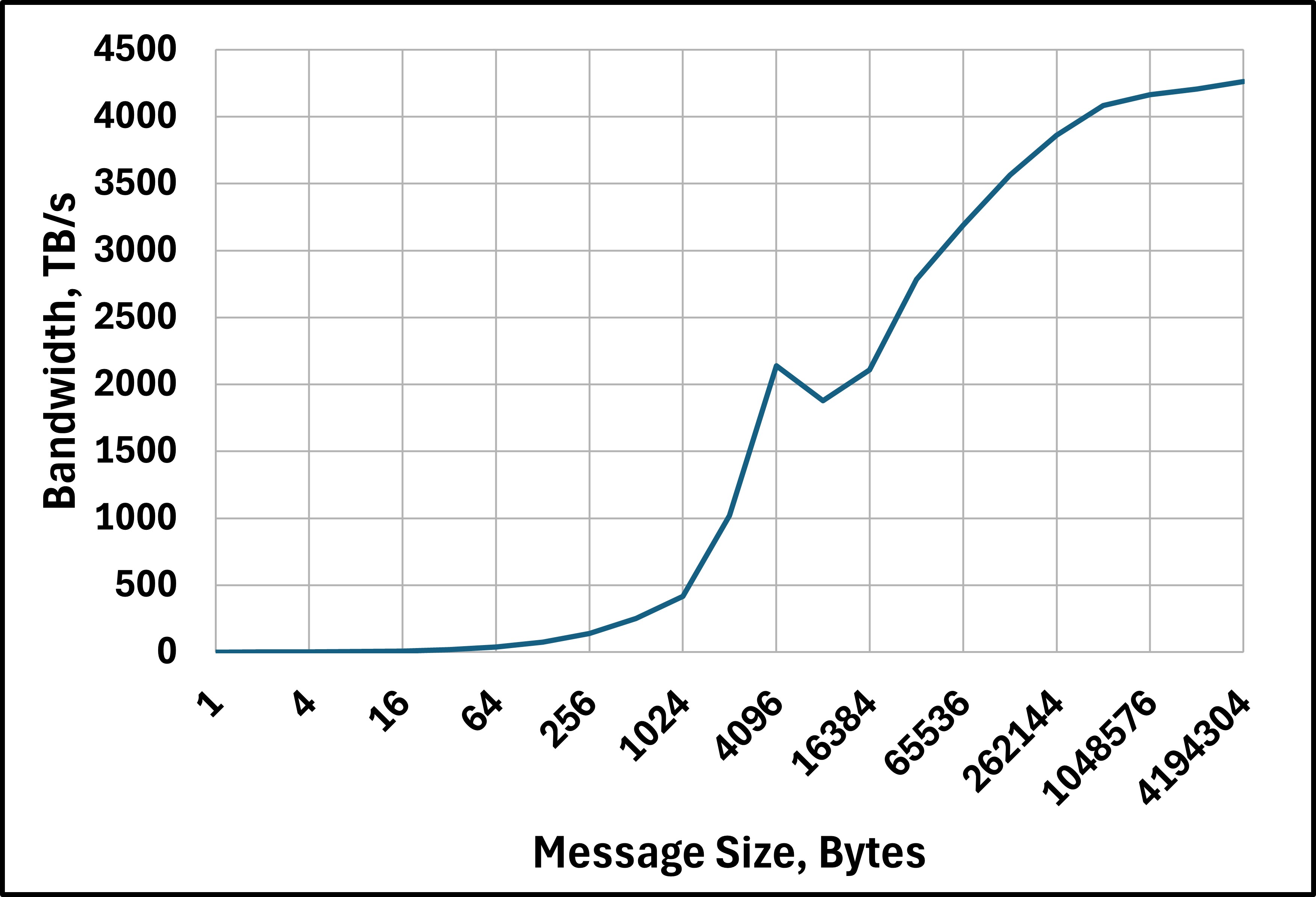}
     \caption{Fabric validation using osu\_mbw\_mr benchmark across 10,262 nodes (82,096 NICs, 41,048 pairs) with 8 processes per node (PPN=8).}
     \label{fig:OSU_Scale_10262}
     \Description{Graph with OSU with 10262 nodes and 82096 NICs.}
 \end{figure}

 Figure~\ref{fig:OSU_Scale_10262} shows the result of the osu\_mbw\_mr  benchmark at a scale of 10262 nodes which is 96.5 percent of the overall system. The system and the nodes were subsequently used to execute several HPC and AI applications that are further explained in \ref{Performance} section.

\subsubsection{CPU Binding and Ranks}
It is important to have the right CPU bindings while executing benchmarks. CASSINI devices can be associated with any of the numa nodes.
The NUMA configuration for Aurora compute host is shown below.

  \begin{itemize}
      \item NUMA node0 CPU(s):     0-51,104-155
      \item NUMA node1 CPU(s):     52-103,156-207
      \item CASSINI devices cxi0-cxi3 are associated with NUMA node 0.
      \item  CASSINI devices cxi4-cxi7 are associated with NUMA node 1. 
\end{itemize}
  
\begin{figure}[h]
    \centering
     \includegraphics[width=0.9\linewidth]{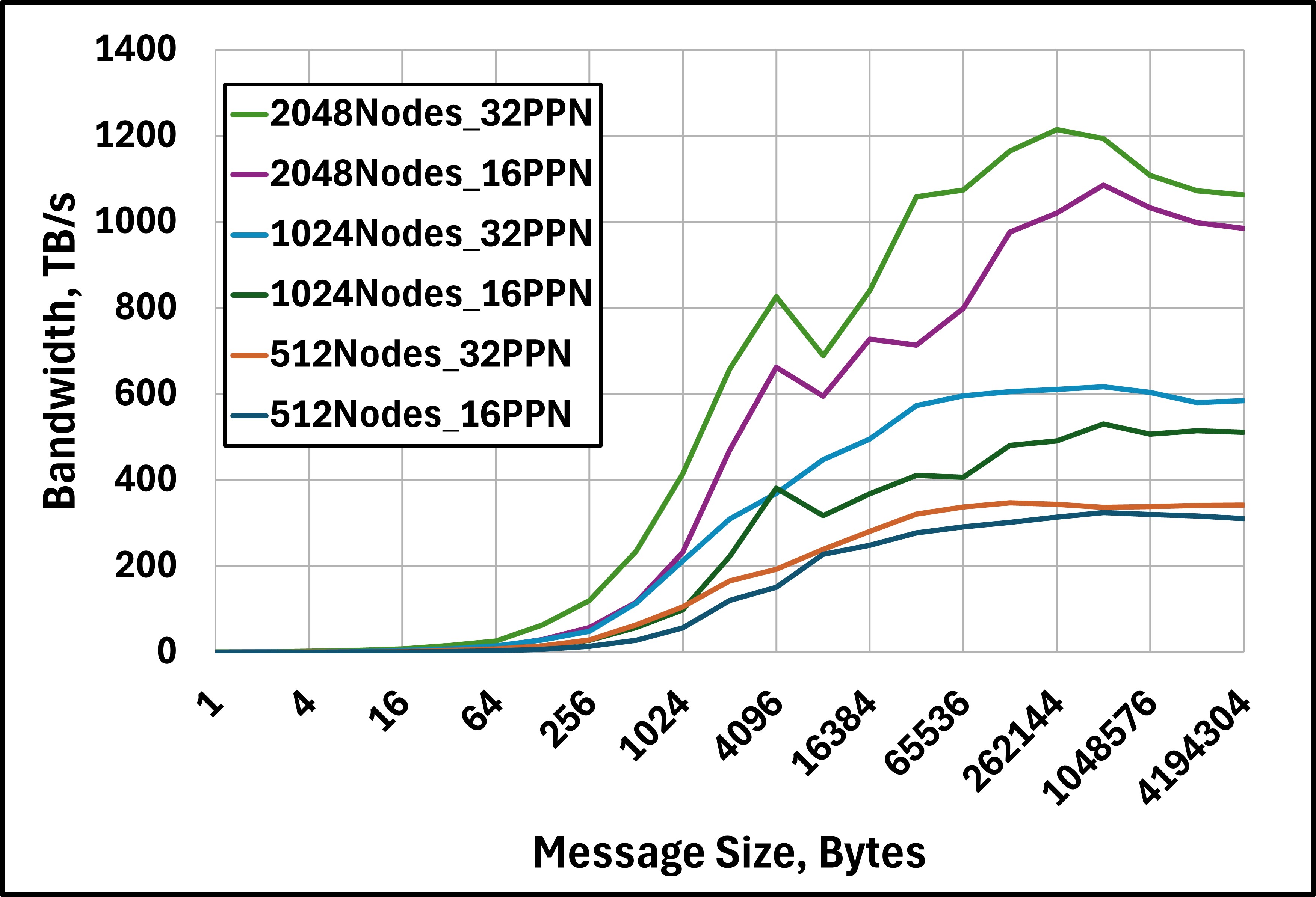}
     \caption{Fabric validation using the osu\_mbw\_mr benchmark across different node counts and PPN.}
     \label{fig:osu_nodes_ppn}
     \Description{The aggregate bandwidth of osu_mbw_mr results demonstrating scaling with higher nodes and higher PPN.}
 \end{figure}

The cpu-bind (CPU binding) option of mpiexec command is used to specifically bind the ranks to the CPU associated with the CASSINI device while executing the network validation benchmarks.  Figure~\ref{fig:osu_nodes_ppn} shows result of osu\_mbw\_mr scaling the number of nodes and also number of processes per node (PPN). A sample script that produces CPU bindings for different PPNs is available at~\cite{cpubinding}.

\subsubsection{Systematic validation}
A systematic approach to fabric validation is used as an effective strategy for validation of the entire system. This involves running fabric validation benchmarks at various levels as listed below.
\begin{itemize}
    \item Node (loopback)
    \item Switch 
    \item Group
    \item System
\end{itemize}

 \subsubsection{Network timeouts}
 Network timeouts are an indication of events (fabric/node) that  require Cassini retries. The Cray MPI library can report the total number of CXI level timeouts during the tests. The line below shows an example timeout summary after results.
 
 \textit{MPICH Slingshot Network Summary: 28 network timeouts.}

 These timeouts require further analysis to understand the cause and potential fix for the same. Slingshot Fabric Monitoring and diagnostic tools are used to correlate events and fix any hardware or software issues.
 Key points to consider are the following.
\begin{itemize}
    \item Network timeouts can be a result of Fabric events
    \item Network timeouts can also be due to node Hardware issues (PCIe, Memory and CPU)
\end{itemize}
Hence the recommended approach is to isolate Fabric and non-Fabric issues to root cause the anomaly.
\subsubsection{Identification of low performing nodes} \label{lpn}
There are several factors that can influence the network performance of a node. It is important to seperate node level issues from fabric level issues.  Node errors are typically logged in console/system logs and hence monitoring the logs with markers before and after the run to identify critical errors helps to isolate low performing nodes. These nodes undergo corrective hardware actions and revalidation. After successful revalidation they are subsequently used for running jobs and applications. Node errors can include:
\begin{itemize}
    \item Hardware errors (PCIe, Memory, CPU, NIC)
    \item CASSINI Edge Link flaps ( A link can have hardware transient errors that can cause the link to be reset and then come back. This can take up ~3-5 seconds for the link to tune and become operational.)
    
\end{itemize}
\subsubsection{CXI Counters}
Performance characterization and insights involve analyzing counters at the host level. HPE Cray MPI offers a feature to automatically gather and analyze Cassini counters for any MPI application run on the HPE Slingshot 11 network. This easy-to-use feature provides important feedback regarding application performance and requires no source code or linking changes. To gather CXI counters for an MPI job,  the following environment variables are used at runtime.
\begin{itemize}
\item MPICH\_OFI\_CXI\_COUNTER\_REPORT
\item MPICH\_OFI\_CXI\_COUNTER\_REPORT\_FILE 
\item MPICH\_OFI\_CXI\_COUNTER\_VERBOSE 
\item MPICH\_OFI\_CXI\_COUNTER\_FILE 
\end{itemize}

Refer to following guides for additional details 
\begin{itemize}
\item HPE Cray Programming Environment guide.
\item HPE Cray Cassini Performance Counters User Guide.
\end{itemize}
\subsubsection{Prolog and Epilog Tests}
Job schedulers and workload managers like SLURM and PBS provide the system administrators the ability to run pre-defined programs (typically scripts) before and after the execution of application. This framework is used to do the require pre-screening before and after scheduling the job. Typical screening tests performed are listed below. 
\begin{itemize}
    \item Prolog tests
    \begin{itemize}
        \item cxi\_healthcheck - Diagnostics that checks health of CASSINI device
        \item cxi\_gpu\_loopback - Tool to check the loopback performance using CASSINI
        \item slingshot-diag - Additional software and hardware diagnostics on CASSINI 
    \end{itemize}
    \item Epilog tests
    \begin{itemize}
        \item CASSINI Flaps - Offline nodes that resulted in CASSINI flaps for hardware action
        \item CASSINI Services Cleanup - Ensure CASSINI services are cleared after any job.
        \item Node Hardware Errors - Offline nodes exceeding thresholds for hardware errors for additional diagnostics.
    \end{itemize}
\end{itemize}

\section{Software}\label{Software}
There are several critical software components that contribute to the successful execution of applications at scale on Aurora.  Fabric management is used to configure, initiaize, and monitor the network.  On the host, the Cassini driver operates the NIC device.  Libfabric provides the low-level communication API for the network. MPI is layered on top of libfabric to provide the familiar application API to the high-speed network.  These software components are described in more detail in this section.

\subsection{Fabric Management Software}\label{Fabric_Management}
\begin{figure*}[htb]
    \centering
     \includegraphics[width=0.80\linewidth]{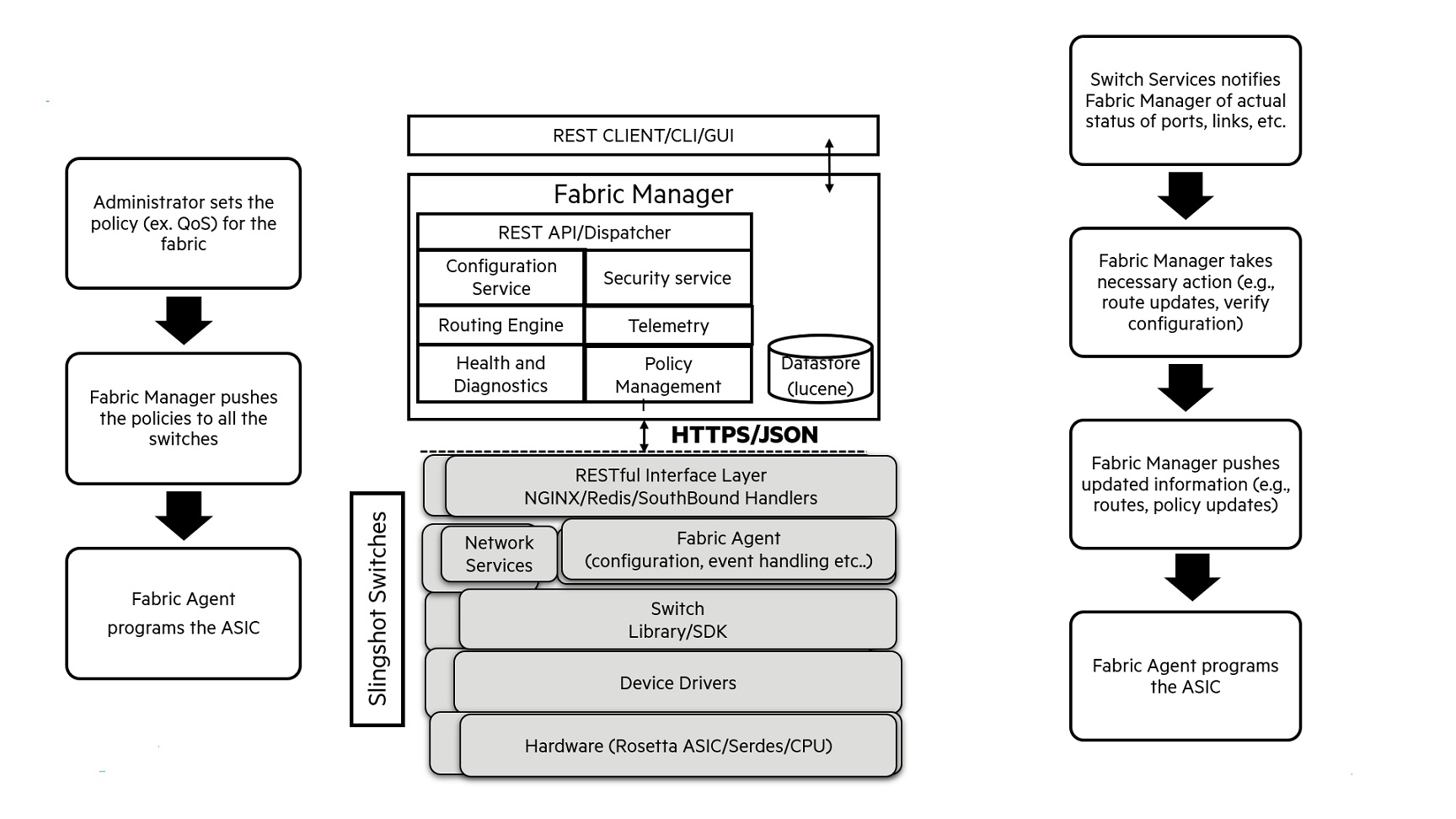}
     \caption{Slingshot Fabric Management.}
     \label{fig:Slingshot_Fabric_Management}
     \Description{Slingshot Fabric Management.}
 \end{figure*}

The HPE Slingshot Fabric Manager includes a suite of software which configures, manages, and monitors the network. The Fabric Manager  runs on an external server and communicates with the switches over the out-of-band management network.
Figure ~\ref{fig:Slingshot_Fabric_Management} describes at a high level the various services running within the fabric manager and the flow of configurations and interaction between Fabric Manager and the switch agents running on the switches. A simulation framework was developed to simulate Slingshot Fabric control path
operation(s) using an unmodified Fabric Manager, Fabric Agent, and a large set of Rosetta switch libraries that are used for configuring and managing a Slingshot Fabric. This enabled HPE to validate that the Fabric Manager is able to scale and manage 5600 switches.

 \subsection{Fabric Configuration}
The section describes some of the important settings that were applied on the Fabric Manager for better scalability and efficiency

\subsubsection{Routing Bias and Group Load Setting}
In Figure~\ref{fig:Aurora_Fabric}, there are 166 compute groups and 8 I/O groups. A group load setting is enabled on the I/O groups to get better-balanced read and write performance. In the absence of software enabling group load to choose lightly loaded intermediate groups, the packets routing non-minimally will follow a more probabilistic model when choosing intermediate groups. By using the software enabled group load setting, adaptive routing is able to use the lightly loaded intermediate group. This ensures that non-minimal adaptive routing is able to optimally utilize all the available global links and intermediate groups when there are read requests from the clients (ordered Puts from I/O groups to a compute group). 

\subsubsection{Fabric Manager Sweep Interval Settings}
There are services in the Fabric Manager that have maintenance routines that occur at predefined cadences. These periodic sweep intervals can be fine-tuned or changed for optimization. The sweep settings must be set appropriately to ensure that the changes in the system are handled effectively. In large systems, setting sweep intervals very aggressively (too low) can increase the load on the Fabric Manager node. At the same time, setting a large value will result in delays in handling the system events and can cause performance impacts. The sweep intervals of the following services can be adjusted for optimization. The default settings used are given below.
\begin{itemize}
    \item Deployment (Default: 10 seconds)
    \item Dragonfly Routing (Default: 5 seconds)
    \item Live Topology (10 seconds)
\end{itemize}

\subsubsection{QoS policies}
Quality of Service (QoS) is used to control traffic and help ensure the performance of applications. The Slingshot Fabric Manager is configured to use a QoS profile that aids in achieving optimal performance.
The profile used is the LlBeBdEt (Profile 2) which creates three bidirectional HPC traffic classes and one dedicated Ethernet traffic class: 
\begin{itemize}
    \item HPC low latency 
    \item HPC bulk data 
    \item HPC best effort 
    \item Ethernet
\end{itemize}
The testing described in this paper only used the HPC Best Effort traffic class for MPI (and Ethernet for IP traffic).

\subsubsection{Orchestrated Maintenance}

The Slingshot Fabric Manager provides orchestrated Maintenance functionality that enables it to perform system diagnostic and maintenance tasks while maintaining the functionality of a running fabric.

Any fabric links identified as problematic can be put to maintenance mode for subsequent diagnosis. This will avoid any impact due to flaps and ensure optimal performance and prevent associated performance anomalies.  This technique is successfully used during HPL and HPL-MxP trial runs to void any performance impact that can manifest as stalls in the application. 

\subsection{Fabric Monitoring}
Effective fabric monitoring is required throughout the lifecycle of the system.
Fabric Monitoring software enables to identify components that are degraded in health and requires diagnosis and maintenance. Fabric Links (local and global) play a significant role in ensuring the optimal performance for the benchmarks and applications. Fabric monitoring enables identification of local and global links that are unhealthy and require hardware action. Monitoring identifies switches that exhibit any hardware errors and required hardware actions. Insights into performance anomalies require analysis of counters from different subsystems. For
example, a system like Aurora  requires more than 300,000 components to be actively monitored. Fabric monitoring software is installed on a separate fabric monitoring node. The fabric monitoring node is an Intel
Xeon Gold 5320 CPU with 256GB RAM with 11TB of storage.

\subsection{Slingshot Host Software Stack}
\begin{figure*}[htb]
    \centering
     \includegraphics[width=0.48\linewidth]{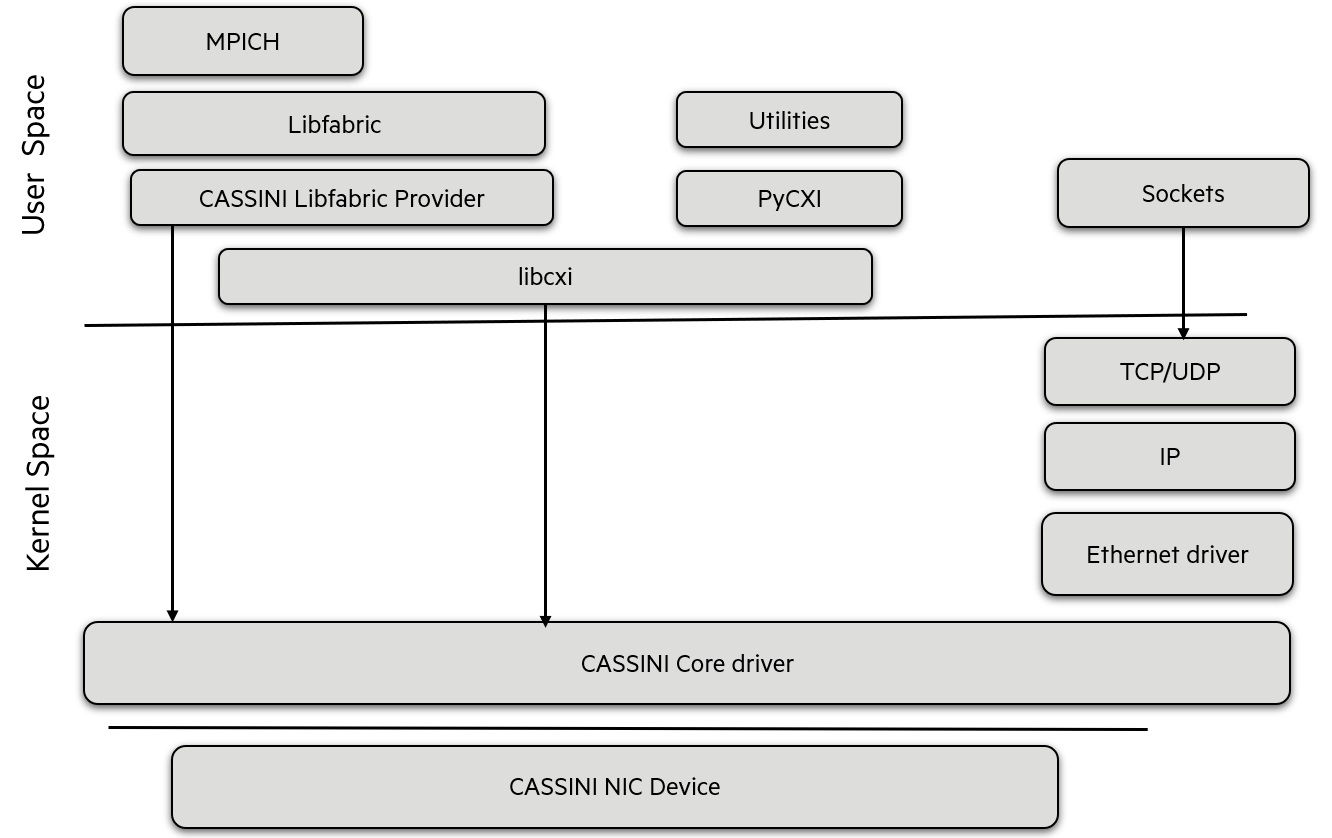}
     \caption{Slingshot Host Software Stack.}
     \label{fig:Slingshot_Host_Software}
     \Description{Slingshot Host Software.}
 \end{figure*}

A high level conceptual overview of Slingshot Host Software stack is shown in figure \ref{fig:Slingshot_Host_Software}. 

 \subsubsection{ Libfabric}
 Libfabric \cite{libfabric} is a high-performance fabric software library designed to provide low-latency interfaces to fabric hardware. Libfabric (OpenFabric Working Group, 2018) provides a set of APIs that match the needs of HPC programming models. They can be implemented in software over hardware interfaces such as verbs or the hardware can implement up to the software interface. Cassini does the latter, providing direct hardware support for many of the libfabric interfaces. MPI implementations including MPICH that is used on Aurora layers on top of libfabric. The CASSINI Libfabric provider enables libfabric on HPE Slingshot Fabric and supports reliable, connection-less endpoint semantics. It supports two-sided messaging interfaces with message matching offloaded by the Cassini NIC. It also supports one-sided RMA and AMO interfaces, light-weight counting events, triggered operations (via the deferred work API), and fabric-accelerated small reductions

 \subsection{MPI Library}
 Aurora's primary MPI library comes from the open source MPICH implementation~\cite{mpich} which contains several optimizations~\cite{PreparingMPICHforExascale} that were needed to overcome the challenges seen in Exascale systems. The MPICH library contains support for communication primitives using GPU buffers, specific to Aurora. Optimizations range from pipelining to host, IPC handles-based exchange for intra-node GPU communication, and improvements to the data type engine. MPICH on Aurora is configured to use the CH4 device that interfaces with libfabric for accessing the Cassini NIC and the Level Zero APIs for the Intel GPUs.

\section{Performance}\label{Performance}

\subsection{MPI Benchmark Performance} \label{performance}

Essential network characteristics such as achieved latencies and bandwidth can be measured at the MPI level using various MPI benchmarks. These quantities provide real world measured values that can be used to evaluate the state of the network, to set application performance expectations, and to interpret achieved application scaling results. Argonne systems have typically been bench-marked using the in-house ALCF MPI benchmark suite \cite{ALCF-MPI-Benchmarks-12} which contains many of the communication patterns typical of ALCF workloads. 

\begin{figure}[htb]
    \centering
     \includegraphics[width=0.9\linewidth]{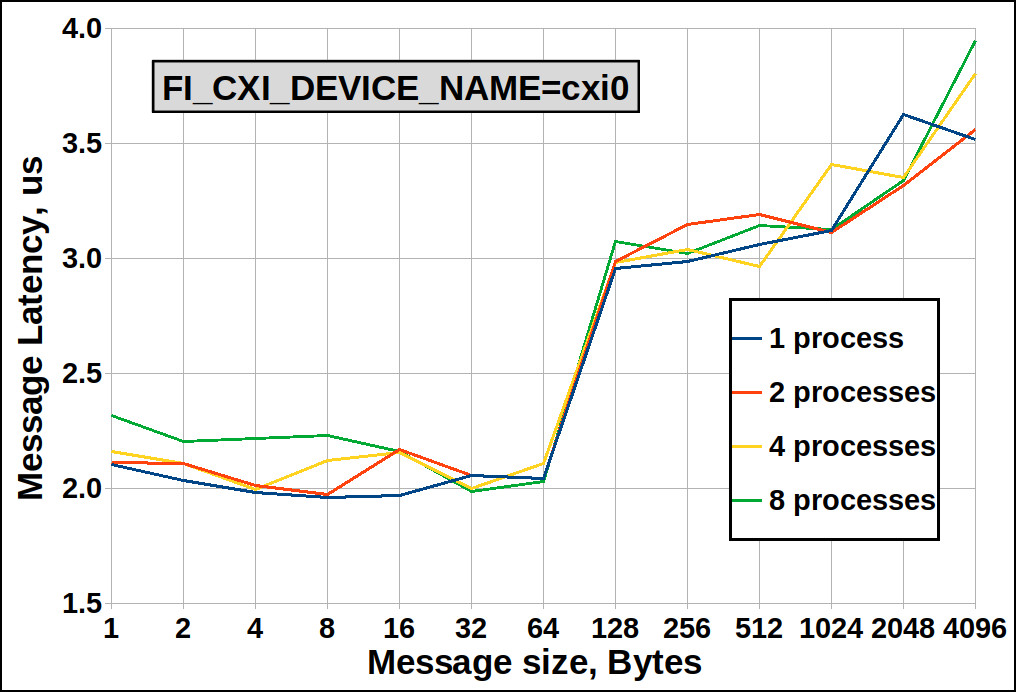}
     \caption{Latency of point-to-point communication operations for different message sizes. }
     \label{fig:MPI_Host_Single_NIC}
     \Description{Graph presents the latency of point-to-point communication operations for different message sizes. Each message is located in a memory of the host. The MPI processes are bound to a single NIC. The latency stays constant for small messages, showing that a NIC can effectively multiplex with little penalty. It would be good to clarify the 1 us latency jump going from 64 Bytes to 128 Bytes. }
 \end{figure}

The latency of point to point operations is a critical performance parameter and is show in Figure \ref{fig:MPI_Host_Single_NIC}. The graph presents the latency of point-to-point communication operations for different message sizes. Each message is located in a memory of the host. The MPI processes are bound to a single NIC. In this particular benchmark we are using synchronous send-recv pairs, and the reported latency is an average over 16 outstanding messages in a communication window. The latency stays constant for small messages, showing that a NIC can effectively multiplex with little penalty. A jump from 64 Bytes to 128 Bytes clearly shows the boundary when message buffering start using host DRAM instead of NIC's SRAM.

\begin{figure}[htb]
    \centering
     \includegraphics[width=0.9\linewidth]{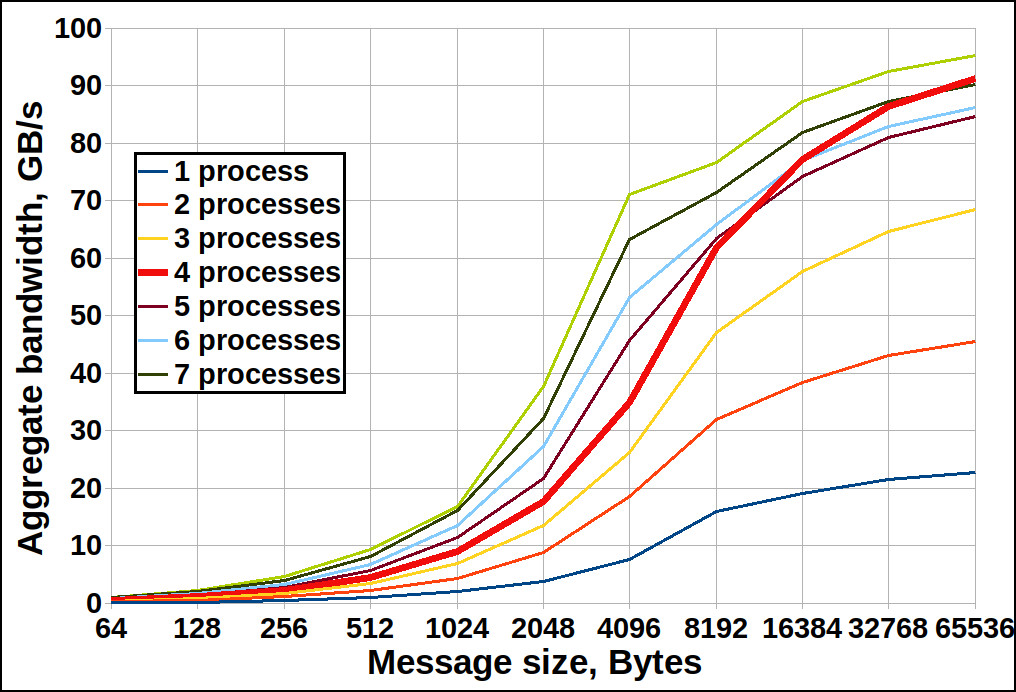}
     \caption{Aggregate off-socket bandwidth for point-to-point communication operations. }
     \label{fig:MPI_Host_Single_socket_4_NICSs}
     \Description{The aggregate off-socket bandwidth for point-to-point communication operations. Multiple MPI processes are placed to the same socket and assigned a NIC in round-robin. The bandwidth is linearly increasing as the processes are added until the point when the processes start sharing the NICs. The red solid line represents the no-sharing one-process-per-NIC scenario with all available NICs utilized. The NICs cannot be saturated by one process and adding the second process adds noticeably to achieve higher bandwidth as one may see from the green line, presenting 8 MPI processes per socket or 2 MPI processes sharing each available NIC.}
 \end{figure}

Another important performance parameter is the achievable off node bandwidth, which is shown in \ref{fig:MPI_Host_Single_socket_4_NICSs}. For this test multiple MPI processes are placed on a single socket and assigned to the four NICs attached to that socket in round-robin fashion. It can be seen that the bandwidth is linearly increasing as the processes are added up to 4 ranks beyond which the processes start sharing the NICs. The red solid line represents the one-process-per-NIC scenario with all available NICs utilized. The NICs cannot be saturated by one process per NIC and adding the second process to each NIC achieve significantly higher bandwidth as shown by the green line, representing 8 MPI processes per socket or 2 MPI processes sharing each available NIC.

\begin{figure}[htb]
    \centering
     \includegraphics[width=0.9\linewidth]{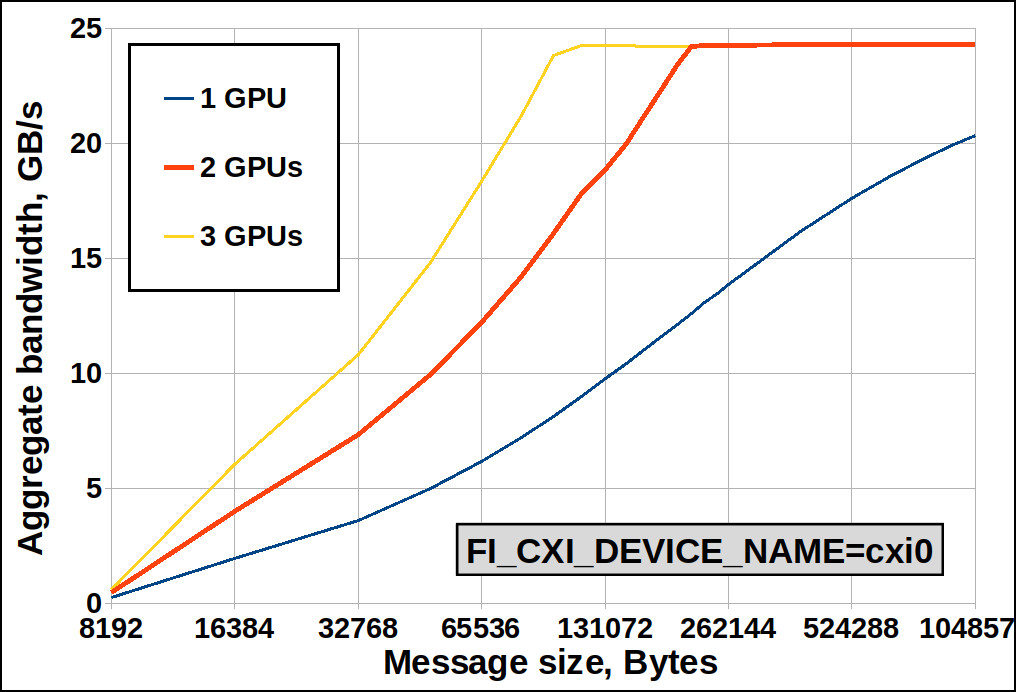}
     \caption{Bandwidth for point-to-point communication operations with buffers located in GPU memory.}
     \label{fig:MPI_Device_Single_NIC}
     \Description{Bandwidth for point-to-point communication operations with buffers located in GPU memory.}
 \end{figure}

\begin{figure}[htb]
    \centering
     \includegraphics[width=0.9\linewidth]{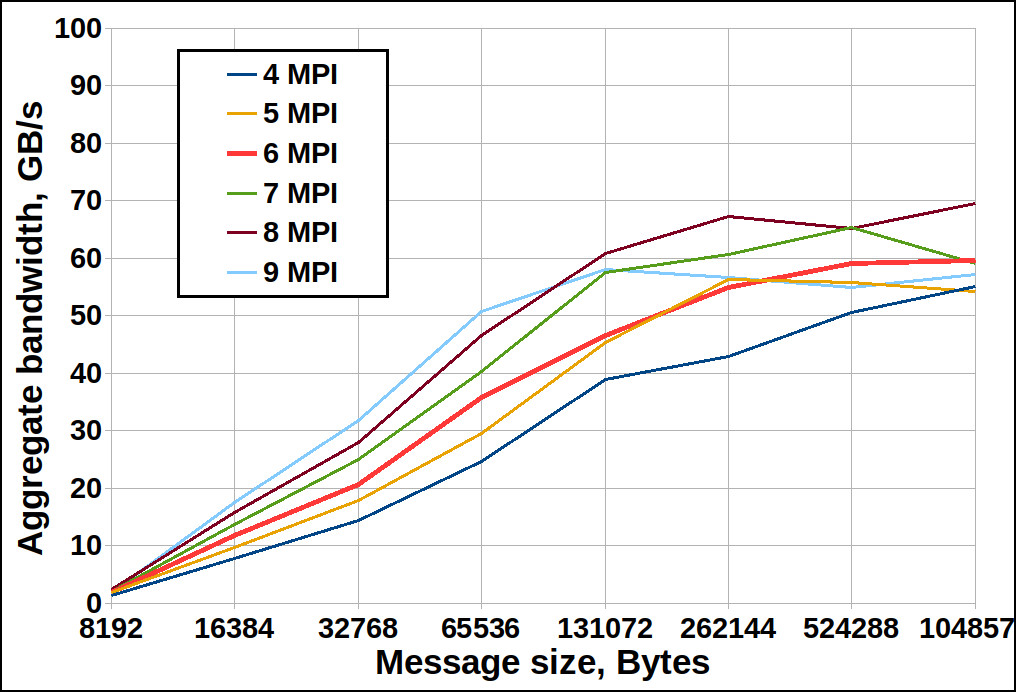}
     \caption{Single socket aggregate bandwidth for point-to-point communication operations with buffers located in GPU memory.}
     \label{fig:MPI_Device_Single_socket_4_NICs}
     \Description{Bandwidth for point-to-point communication operations with buffers located in GPU memory. 4 NICS can give no more than 70~GB/s aggregate bandwidth, while if the buffers located on host memory, the effective bandwidth reaches 90~GB/s.}
 \end{figure}

MPI on Aurora supports moving data directly to and from GPU memory and Figure \ref{fig:MPI_Device_Single_NIC} shows bandwidth for point-to-point communication operations with buffers located in GPU memory. In this test, each MPI process is bound to a different GPU and the same single NIC. The plot shows that a single process cannot saturate the NIC's bandwidth for message sizes up to a megabyte. Adding additional processes allows reaching an effective bandwidth of 23~GB/s at around 256~KB per message. Figure \ref{fig:MPI_Device_Single_socket_4_NICs} show single socket aggregate bandwidth across all available network cards for point-to-point communication operations with buffers located in GPU memory. Each MPI process is bound to a different GPU and a different NIC, both bindings are assigned in round-robin fashion. When running four MPI processes on a socket, one pair of processes shares the same NIC. According to our measurements, balancing the NIC assignments is a key to reach the effective aggregate bandwidth, which is around 70~GB/s. We attribute the loss of performance to PCIe Gen 4 - PCIe Gen 5 conversion inefficiencies since with buffers in the host 90 GB/s is achievable. 

\begin{figure}[htb]
    \centering
     \includegraphics[width=0.9\linewidth]{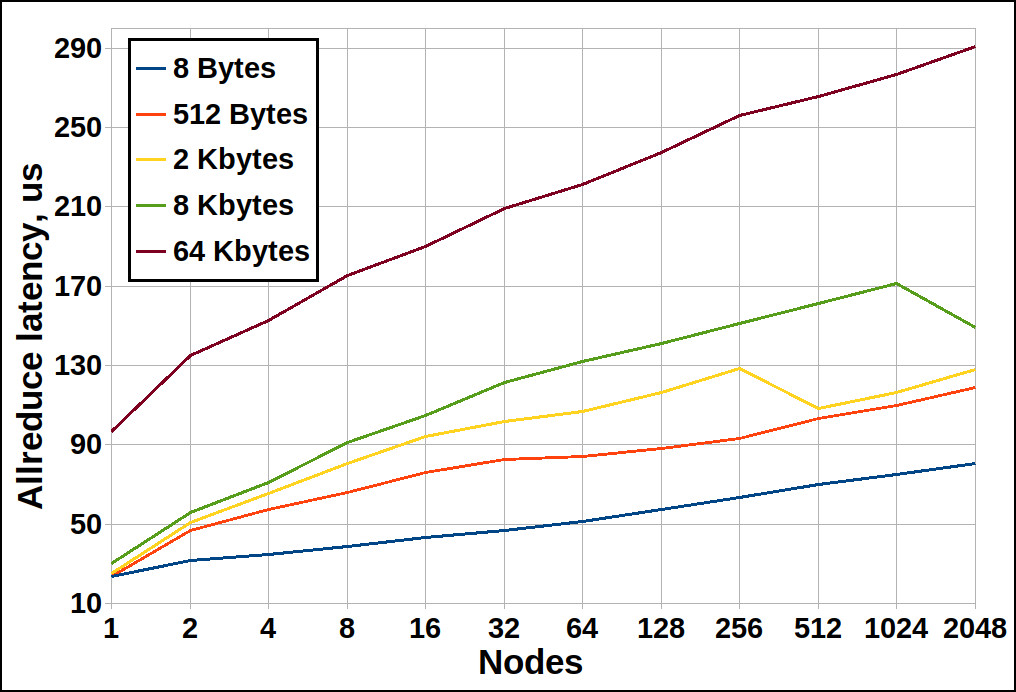}
     \caption{Latency for MPI reduction operation for buffers located in GPU memory. }
     \label{fig:MPI_Allreduce}
     \Description{Latency for MPI_Allreduce.}
 \end{figure}

An important element of applications performance at larger scales is the cost of MPI collective operations. MPI provides a number of different collective calls but a frequently utilized one is {\sf MPI\_Allreduce}. Figure \ref{fig:MPI_Allreduce} shows {\sf MPI\_Allreduce} call performance on Aurora at different node counts up to 2048 nodes and for different message sizes. Less than linear latency growth is observed, which is typical for a recursive-doubling tree algorithm. A switch from a ring algorithm to a tree algorithm is clearly seen on the curves.

\subsection{Scalable Benchmark Performance} \label{performance}

The results of four benchmarks presented in this section. HPL (High Performance Linpack) is the standard benchmark used to generate the Top500 list, emphasizing 64-bit floating-point performance. HPL-MxP is a mixed-precision variant of HPL designed to better represent the performance of systems on AI-oriented workloads. Graph500 targets data-intensive workloads and emphasizes memory access patterns and graph traversal rather than floating-point operations. HPCG (High Performance Conjugate Gradients) is also a computational benchmark but focuses on memory bandwidth, data access patterns, and communication, providing a more balanced assessment of system performance. Details of the execution of each benchmark on the Aurora system are presented below.

\subsubsection{HPL}\label{HPL}
\begin{figure}[htb]
    \centering
     \includegraphics[width=0.9\linewidth]{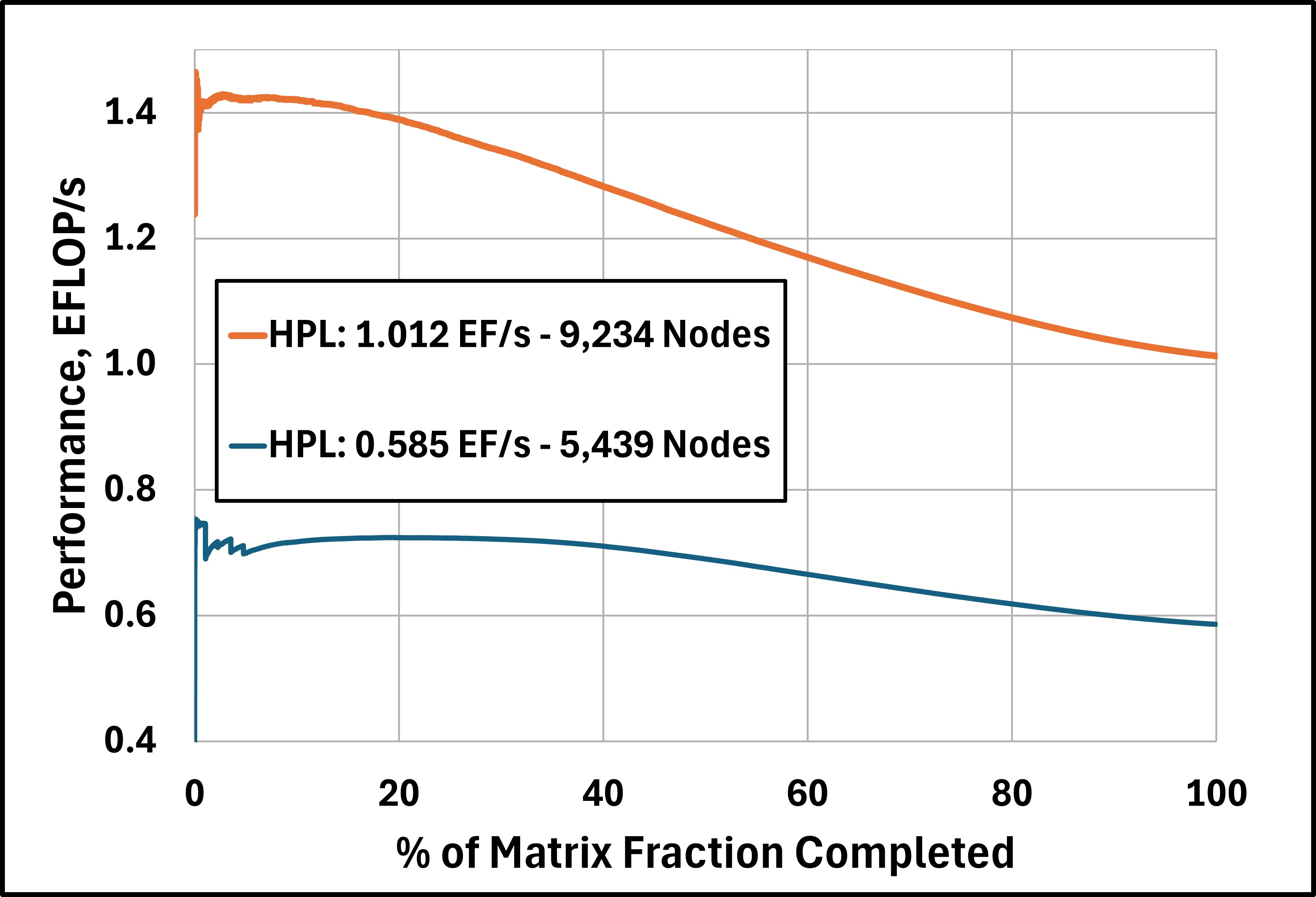}
     \caption{Performance of HPL on 5,439 and 9,234 nodes.}
     \label{fig:HPL_Performance}
     \Description{Graph with HPL performance (X-axis) as a function of the fraction completion (X-axis).}
 \end{figure}

On the HPL benchmark Aurora achieved a performance of 1.012~EF/s using 9,234 nodes (approximately 87\% of the full system), ranking third on the Top500 list at the time of submission to SC24~\cite{TOP500}. The MPI process grid was configured with \( P = 162 \) and \( Q = 342 \), where \( P \) and \( Q \) represent the number of MPI processes along each dimension. This configuration resulted in a scaling efficiency of 78.84\% over a runtime of 4 hours, 21 minutes, and 54 seconds, where scaling efficiency is defined as the ratio of the achieved performance per node to the theoretical peak performance per node. As shown in Figure~\ref{fig:HPL_Performance}, performance remained relatively smooth across all phases of computation, from dense LU factorization to the iterative refinement (IR) phase and final result computation. Minor performance degradation was observed during the initial phase, suggesting an opportunity for further optimization to improve scaling efficiency.

At 5,439 nodes, Aurora achieved a performance of 585~PF/s, as also shown in Figure~\ref{fig:HPL_Performance}. Scaling efficiencies across different node counts are summarized in Table~\ref{tab:HPL_Scaling_Efficiency}.

\begin{table}[h]
  \caption{HPL Scaling Efficiency.}
  \label{tab:HPL_Scaling_Efficiency}
  \begin{tabular}{ccc}
    \toprule
    Number of Nodes&Performance (PF/s)&Scaling Efficiency(\%)\\
    \midrule
    9,234 & 1012 & 78.84\\
    8,748 & 954.43 & 78.49\\
    8,632 & 949.02 & 79.10\\
    8,109 & 873.78 & 77.52\\
    8,058 & 865.93 & 77.31\\
    7,200 & 805.24 & 80.46\\
    6,888 & 764.04 & 79.80\\
    6,273 & 688.99 & 79.02\\
    5,439 & 585.43 & 77.44\\
  \bottomrule
\end{tabular}
\end{table}


\subsubsection{HPL-MxP} \label{HPL-MxP}
\begin{figure}[htb]
    \centering
     \includegraphics[width=0.9\linewidth]{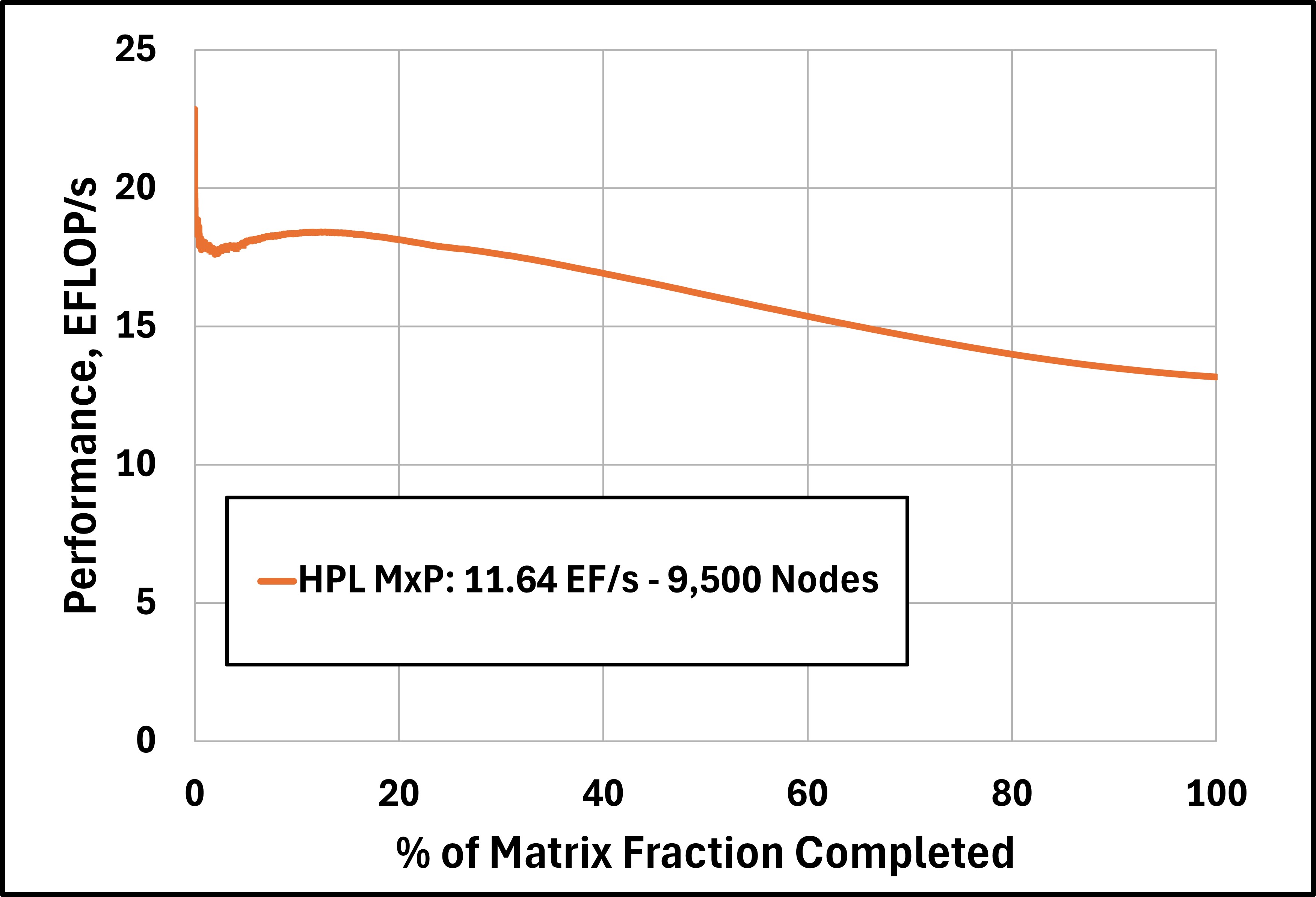}
     \caption{Performance of HPL MxP with 9,500 nodes.}
     \label{fig:HPL_MXP_Performance}
     \Description{Graph with HPL MxP performance (X-axis) as a function of the fraction completion (X-axis).}
 \end{figure}

On HPL-MxP Aurora achieved a performance of 11.64~EF/s using 9,500 nodes (approximately 89\% of the full system), ranking first on the HPL-MxP list at the time of submission to SC24~\cite{HPLMxP}. The LU factorization phase employed mixed-precision arithmetic, utilizing FP16 and FP32 formats, while the iterative refinement (IR) phase uses FP64. As shown in Figure~\ref{fig:HPL_MXP_Performance}, performance scaled uniformly across the phases of computation and communication. Nevertheless, slight performance degradation was observed during the initial and final phases, indicating room for further improvement. Enhancing compute and communication resource utilization, particularly by reducing latencies in broadcast and swap operations, could further improve performance in these regions.

\subsubsection{Graph500}
For the Graph500 BFS benchmark, Aurora achieved a performance of 69,373 GTEPS at scale 42 with 8,192 nodes ( approximately 77\% of the full system)~\cite{Graph500}.

\subsubsection{HPCG}
Aurora's High Performance Conjugate Gradients (HPCG) benchmark achieved a performance of 5.613 PFLOPS with 4,096 nodes (approximately 39\% of the full system), currently ranked third on the HPCG list~\cite{HPCG}. HPCG is complementary to HPL with its computations, data access patterns, and MPI communications as a representative for broad set of HPC applications.

\subsection{Application-Level Performance}

 To evaluate system-level performance across a diverse range of scientific workloads, the scaling and communication of five representative applications were evaluated including: HACC, Nekbone, AMR-Wind, LAMMPS, and FMM. These applications span domains including cosmology, quantum chemistry, molecular dynamics, and computational fluid dynamics, and exhibit a variety of computational characteristics, from dense linear algebra and FFTs to irregular memory access patterns and communication-intensive kernels.
 
\subsubsection{HACC}

The Hardware/Hybrid Accelerated Cosmology Code (HACC) is a high-performance simulation framework designed to solve large-scale cosmological problems efficiently~\cite{HABIB201649}. HACC employs advanced techniques such as Recursive Coordinate Bisection (RCB) trees for particle interactions and domain decomposition. It is widely used for simulating dark matter evolution, galaxy formation, and large-scale cosmic structures.

HACC computations consist of three primary phases: (1) a short-range force evaluation kernel, which is compute-intensive and features regular, stride-one memory access patterns; (2) a tree-walk phase, characterized by irregular indirect memory accesses, frequent branching, and a high volume of integer operations; and (3) a long-range force calculation using 3D FFTs, which is dominated by point-to-point communication.

\begin{figure}[htbp]
\centerline{\includegraphics[width=0.9\linewidth]{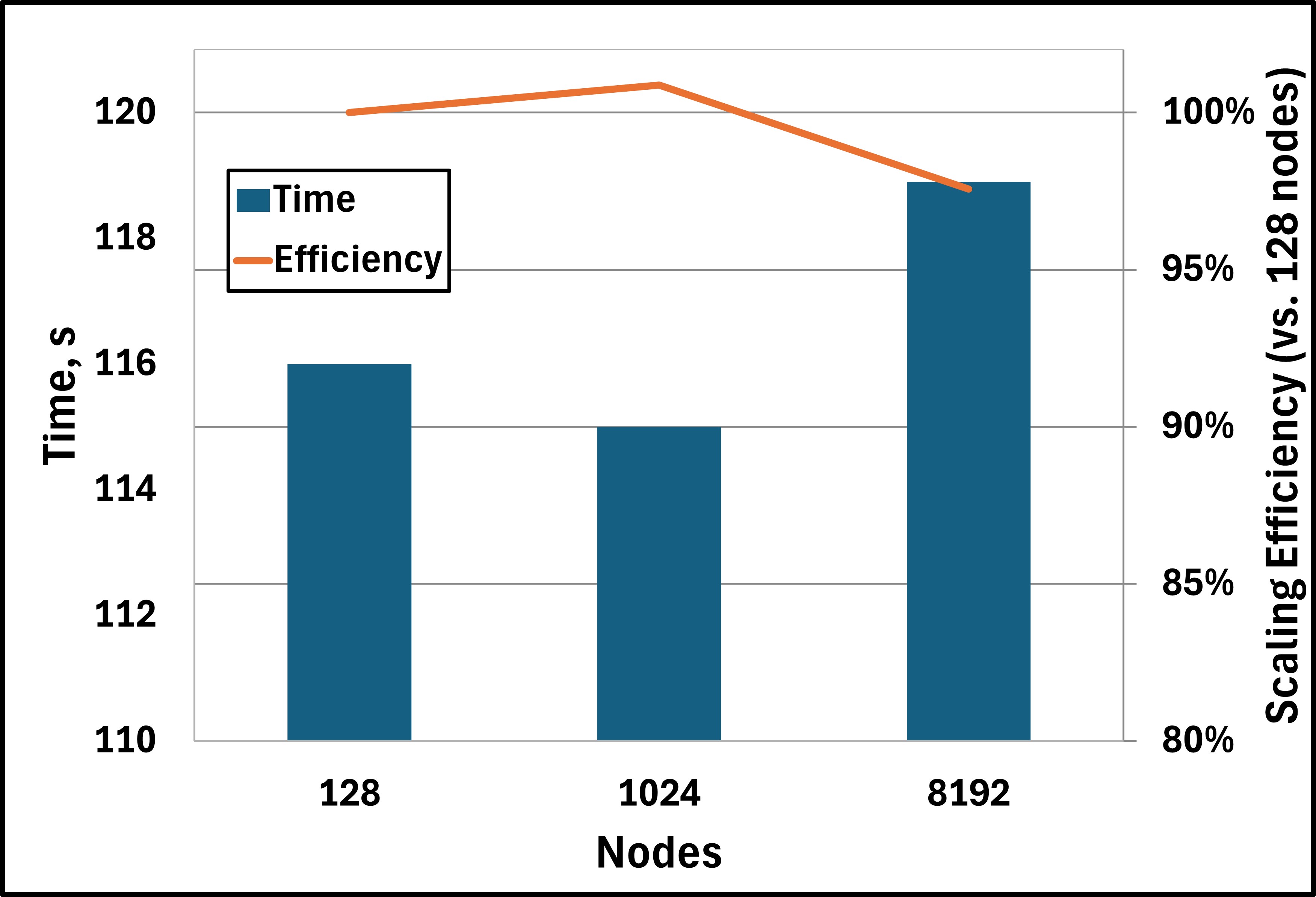}}
\caption{Weak scaling performance and efficiency of HACC. Bars represent execution time (lower is better); lines indicate scaling efficiency.}
\label{fig:hacc_scaling}
\end{figure}

Figure~\ref{fig:hacc_scaling} presents the weak scaling performance and efficiency at 128, 1,024, and 8,192 nodes~\cite{ibeid2025hbm}. Because the simulation domain is cubic, increasing the node count by a factor of eight doubles the grid size (\(n_g\)) along each dimension, resulting in an eightfold increase in the total number of particles. Table~\ref{tab:hacc_config} summarizes the grid sizes and MPI geometries, assuming 96 MPI processes per node (PPN = 96).

\begin{table}[htbp]
    \centering
    \caption{HACC configurations.}
    \renewcommand{\arraystretch}{1.3}
    \begin{tabular}{ccc}
        \toprule
        \textbf{Node Count} & \textbf{Grid Size (\(n_g\))} & \textbf{MPI Geometry} \\ 
        \midrule
         128 &  4608 & $32 \times 24 \times 16$ \\
         1024 &  9216 & $64 \times 48 \times 32$ \\
         8192 & 18432 & $128 \times 96 \times 64$ \\
        \bottomrule
    \end{tabular}
    \label{tab:hacc_config}
\end{table}

As shown in Figure~\ref{fig:hacc_scaling}, HACC maintains excellent weak scaling. Efficiency remains near 99\% at 1,024 nodes relative to the 128-node baseline, and decreases slightly to 97\% at 8,192 nodes, demonstrating strong scalability at extreme scales.

\subsubsection{Nekbone}

Nekbone~\cite{nekbone2015} is a proxy application derived from Nek5000~\cite{fischer2008nek5000}, a high-order spectral element CFD code for solving the incompressible Navier--Stokes equations. Nekbone solves a Poisson problem using an iterative conjugate gradient solver with a simple preconditioner, capturing the essential numerical and communication patterns of Nek5000 while significantly simplifying the workload.

The code is implemented in Fortran and C, with Fortran handling most numerical operations and C handling communication routines. Each conjugate gradient iteration includes vector updates, matrix-matrix multiplications, nearest-neighbor halo exchanges, and global MPI \texttt{Allreduce} operations.

\begin{figure}[htbp]
\centerline{\includegraphics[width=0.9\linewidth]{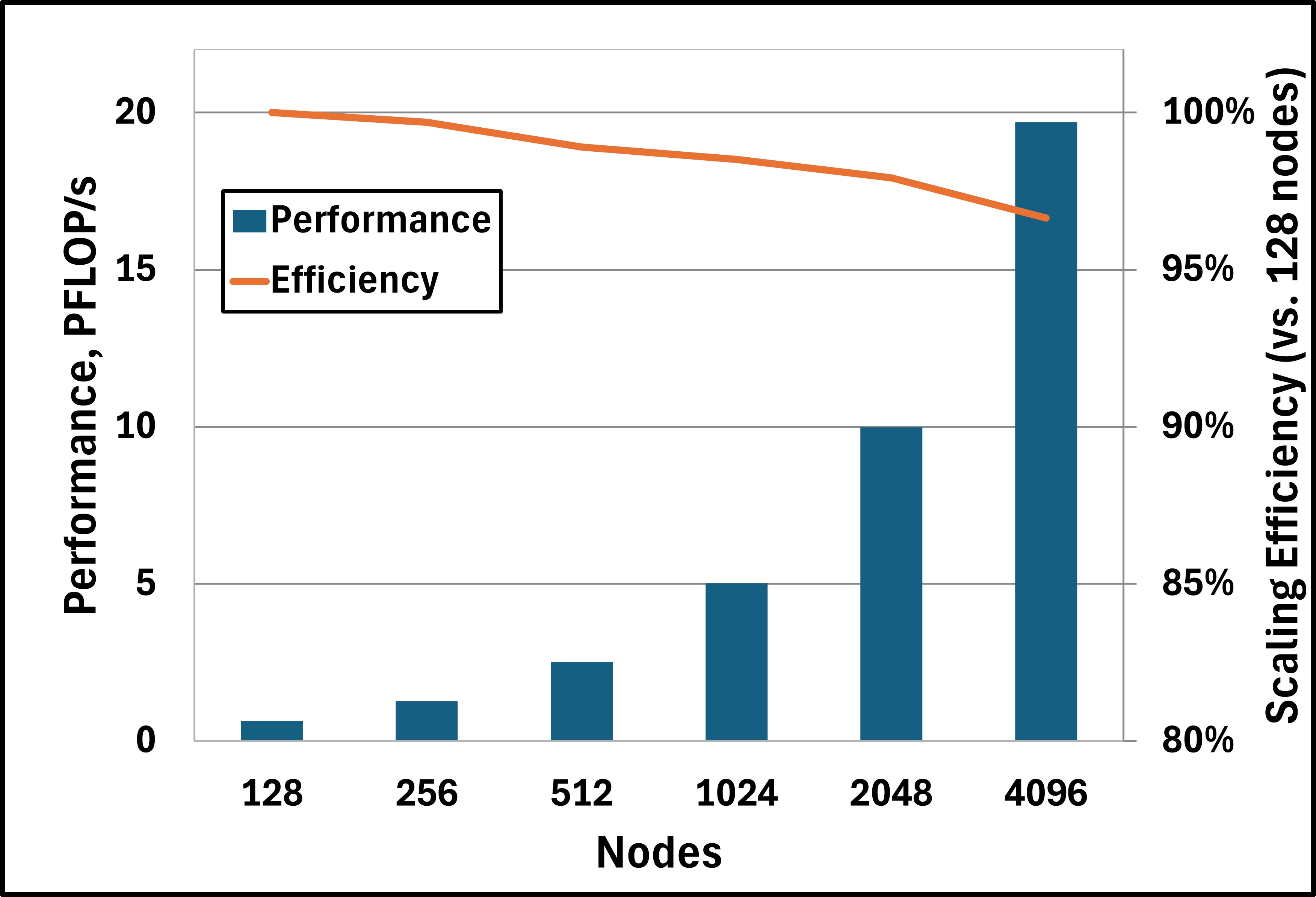}}
\caption{Weak scaling performance and efficiency of Nekbone. Bars represent performance (higher is better); lines indicate scaling efficiency.}
\label{fig:nekbone_scaling}
\end{figure}

The benchmark uses 42,000 spectral elements per MPI process. Two polynomial orders, \(nx1 = 9\) and \(nx1 = 12\), are used to evaluate performance across different resolutions. Figure~\ref{fig:nekbone_scaling} shows the weak scaling performance on Aurora, using 12 MPI processes per node (PPN = 12). Performance is reported as average PFLOP/s across both polynomial orders. Nekbone demonstrates excellent weak scaling with over 95\% parallel efficiency up to 4,096 nodes.

\subsubsection{AMR-Wind}
AMR-Wind~\cite{amrwind2025} is a block-structured adaptive-mesh, incompressible flow solver for wind farm simulations. The primary applications for AMR-Wind are performing large-eddy simulations (LES) of atmospheric boundary layer (ABL) flows, and as a background solver when coupled with a near-body solver with overset methodology to perform blade-resolved simulations of multiple wind turbines within a wind farm. 

AMR-Wind is built on top of the AMReX library~\cite{AMReX_JOSS}, and AMReX provides SYCL backend for Intel GPUs. The codebase is a wind-focused fork of incflo, and the mesh data structures, mesh adaptivity,  as well as the linear solvers for solving the governing equation are provided by AMReX library.  

\begin{figure}[htbp]
\centerline{\includegraphics[width=0.9\linewidth]{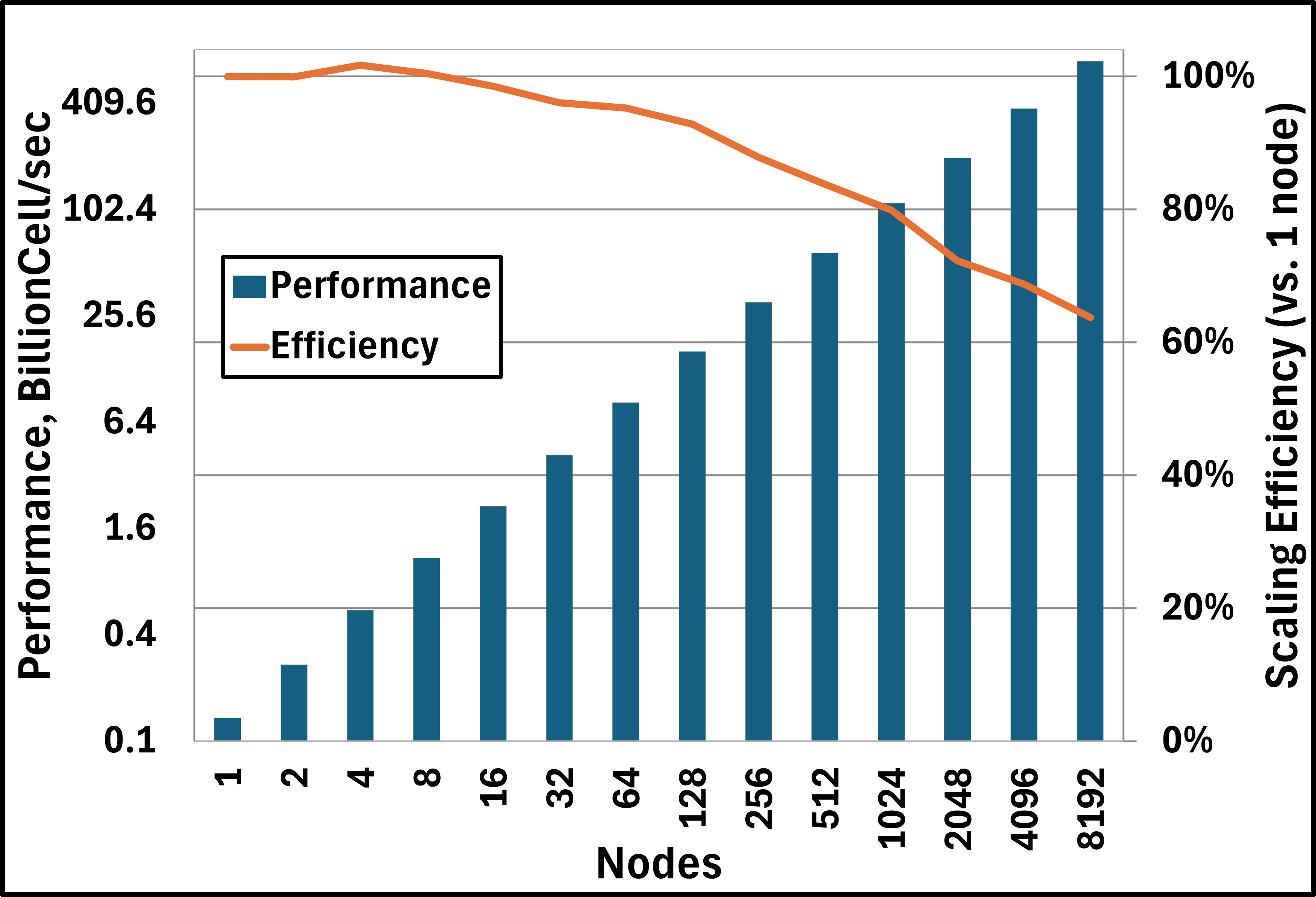}}
\caption{Weak scaling performance and efficiency of AMR-Wind. Bars represent performance (higher is better); lines indicate scaling efficiency.}
\label{fig:AMR-Wind}
\end{figure}

Figure ~\ref{fig:AMR-Wind} shows the weak scaling performance for solving an atmospheric boundary layer flow in a cubic box. The plot shows FOM data defined by a billion cells simulated per second for each time step with up to 8,192 nodes on Aurora. Each MPI rank solves \(256 \times 256 \times 256\) cells for the the \(1{,}000 \times 1{,}000 \times 1{,}000\) domain, and each compute node uses 12 MPI ranks (PPN = 12) to solve \(1{,}024 \times 768 \times 256\) cells for \(4{,}000 \times 3{,}000 \times 1{,}000\) domain. As the number of nodes increases, the domain size is increased in the x and y axes, while the cells and domain along the z axis are fixed as 256 and 1,000 respectively. 

\subsubsection{LAMMPS}
The Large-scale Atomic/Molecular Massively Parallel Simulator (LAMMPS) is a highly scalable molecular dynamics code designed for the simulation of materials and biomolecular systems~\cite{thompson2022lammps}. The LAMMPS Rhodopsin benchmark models an all-atom Rhodopsin protein embedded in a solvated lipid bilayer, employing the CHARMM force field, long-range Coulombic interactions via the particle-particle particle-mesh (PPPM) method, SHAKE constraints, and a reduced water model.

In the largest configuration, the simulation spans 254 billion atoms across 9,216 nodes, using 96 MPI processes per node (PPN = 96) and two OpenMP threads per process. The domain is decomposed into a \(96 \times 96 \times 96\) process grid. Each MPI process further subdivides its subdomain into a spatial binning of \(4 \times 6 \times 4\) to optimize neighbor list construction, a key technique for reducing the cost of pairwise force calculations at extreme scales.

\begin{figure}[htbp]
\centerline{\includegraphics[width=0.9\linewidth]{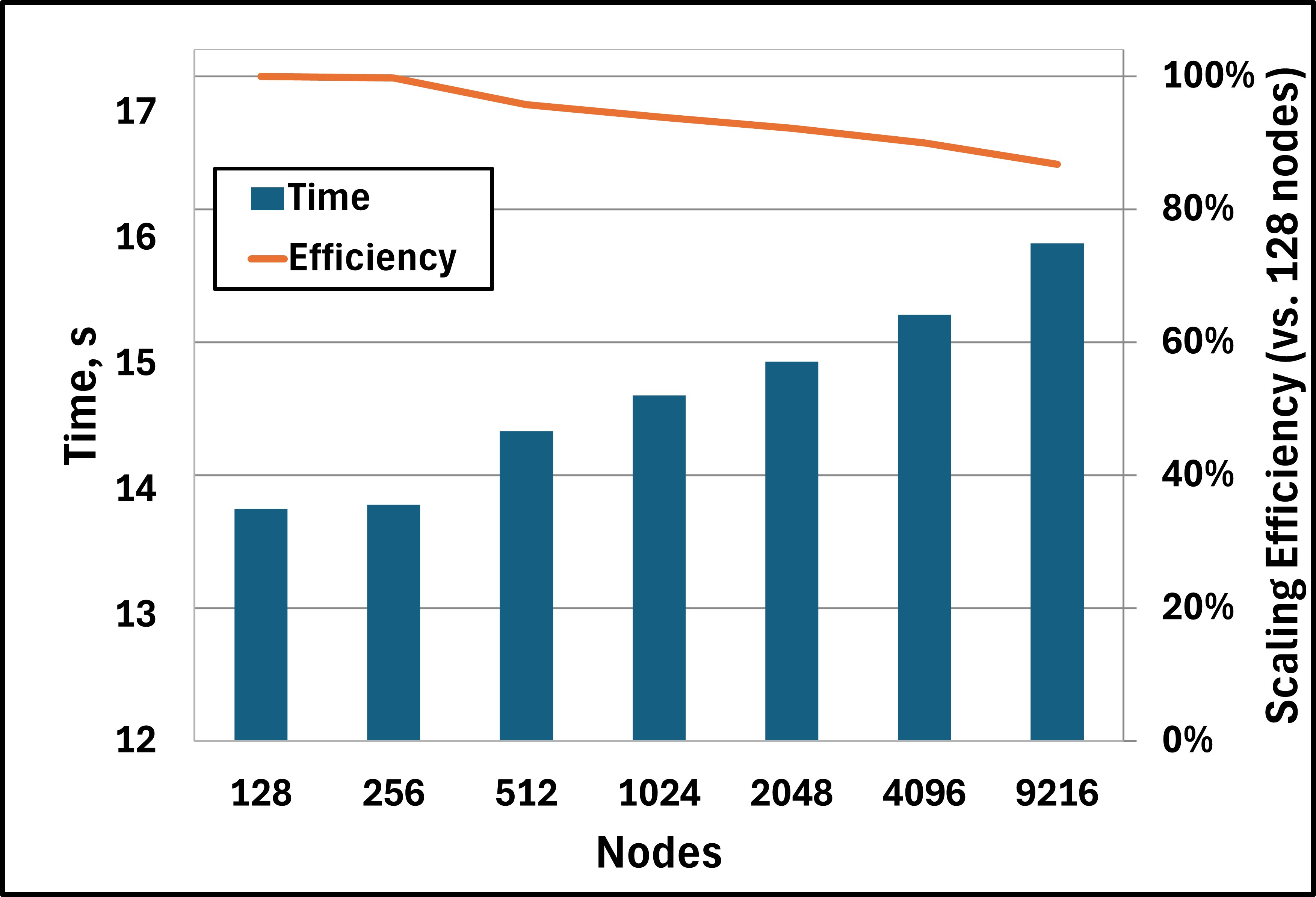}}
\caption{Weak scaling performance and efficiency of LAMMPS. Bars represent execution time (lower is better); lines indicate scaling efficiency.}
\label{fig:lammps_scaling}
\end{figure}

Figure~\ref{fig:lammps_scaling} shows the weak scaling performance as the system scales from 128 to 9,216 nodes. As the node count increases, the total number of atoms scales linearly, with corresponding growth in the physical domain along all three dimensions. The simulation maintains over 85\% parallel efficiency at 9,216 nodes relative to the baseline performance at 128 nodes.

\subsubsection{FMM}

The Fast Multipole Method (FMM) is a part of the electronic structure theory package NWChemEx~\cite{kowalski2021nwchem}. It computes long-range electrostatic interactions between charged particles in a linear-scaling fashion. The FMM application has an irregular communication pattern. In it, each process requests a large number of data pieces sparsely populated in the local memory of numerous remote ranks. Each rank can be a sender and a receiver for every other rank and those roles constantly flip. This communication pattern makes manually arranging synchronization between senders and receivers impractical and therefore a one-sided communication model ideally fits into this pattern. One-sided communication, also known as remote memory access (RMA), is a MPI programming model for data transfer between remote memory locations. The greatest advantage of one-sided communication is that it does not require explicit synchronization between the communicating processes. To start the communication, each process registers a window in their local memory for read or write by remote processes.  The process issues MPI\_Put command to write data to the window on a remote process, or MPI\_Get to read the data from the remote process window. Both commands work asynchronously so there is no need to wait for completion of the data transfer. Closing the RMA window finalizes the one-sided communication between the processes. In FMM each rank issues MPI\_Get requests to a remote rank. The ranks holding the data are unaware and unaffected by the incoming MPI\_Get requests, so they can focus on arranging their own MPI\_Get requests.  As a result, the one-sided communication implementation in the FMM code takes only a dozen of lines of code.

Despite the potential benefits of one-sided communication it has historically been less frequently utilized in HPC than two-sided communication methods. As a result the implementation of one-sided communication is often less tested and less mature, particularly on new platforms. Therefore a set of tests have been developed and utilized to test the functionality and performance of one-sided communication performance on Aurora as part of the development of the FMM code on Aurora. One issue identified in this process is that the PVC GPU has been found to be unable to provide RMA support in hardware and instead the needed functionality has been implemented in software. A partial mitigation is to call MPI\_Win\_fence after certain number of MPI\_Get or MPI\_Put calls to flush the internal buffer. Also, MPICH incorporates environment variable MPIR\_CVAR\_CH4\_OFI\_ENABLE\_HMEM to use HBM memory for RMA that provides a significant performance improvement.

\begin{table}[htbp]
    \centering
    \caption{Configuration of one-sided tests.}
    \renewcommand{\arraystretch}{1.3}
    \begin{tabular}{ccc}
        \toprule
        \textbf{N Nodes} & \textbf{N Particles} & \textbf{N Total Messages} \\ 
        \midrule
         $1 \times 8$ & $1.3 \cdot 10^8$ & 1,615,459 \\
         $1 \times 16$ & $1.3 \cdot 10^8$ & 2,127,199 \\
         $1 \times 32$ & $1.3 \cdot 10^8$ & 2,776,246 \\
         $9 \times 16$ & $1.0 \cdot 10^{11}$ & 19,201.665 \\
        \bottomrule
    \end{tabular}
    \label{tab:onesided_config}
\end{table}

\begin{table}[htbp]
    \centering
    \caption{Time, sec to complete data transfer by using MPI\_Get.}
    \renewcommand{\arraystretch}{1.3}
    \begin{tabular}{ccc}
        \toprule
        \textbf{N Nodes} & \textbf{with HMEM} & \textbf{without HMEM} \\ 
        \midrule
         $1 \times 8$ & 0.9 & 24.6 \\
         $1 \times 16$ & 1.1 & 17.1 \\
         $1 \times 32$ & 1.6 & 13.0 \\
         $9 \times 16$ & 14.5 & NA \\
        \bottomrule
    \end{tabular}
    \label{tab:onesided_get}
\end{table}

\begin{table}[htbp]
    \centering
    \caption{Time, sec to complete data transfer by using MPI\_Put.}
    \renewcommand{\arraystretch}{1.3}
    \begin{tabular}{ccc}
        \toprule
        \textbf{N Nodes} & \textbf{with HMEM} & \textbf{without HMEM} \\ 
        \midrule
         $1 \times 8$ & 14.2 & 28.4 \\
         $1 \times 16$ & 17.6 & 38.9 \\
         $1 \times 32$ & 20.7 & 49.7 \\
        \bottomrule
    \end{tabular}
    \label{tab:onesided_put}
\end{table}

The communication pattern in the FMM code is one-to-all for every rank. It has theoretically a quadratic scaling when increasing the number of ranks so one would want to use the minimum necessary number of ranks in the FMM job just to provide the required aggregate GPU memory for the computation. Tests conducted on Aurora, using the configurations in Table~\ref{tab:onesided_config}, lead to following conclusions. As shown in Tables~\ref{tab:onesided_get} and \ref{tab:onesided_put}, the implementation of MPI\_Get is an order of magnitude faster than that of MPI\_Put. The use of HMEM gives an order of magnitude speed up for MPI\_Get, whereas its benefit for MPI\_Put is considerably lower, only a factor of two. An application using one-sided communications in Aurora must periodically call MPI\_Win\_fence to flush the one-sided buffer and prevent its overflow. The conducted tests worked with the fence being called every 2000 MPI\_Get or MPI\_Put calls with or without HMEM except the case when using MPI\_Put without HMEM where it had to be set to 100 to prevent the communication failure. Using multiple sub-communicators, such as using  nine in the fourth configuration ($9 \times 16$) in comparison to one communicator as shown in the second configuration ($1 \times 16$) while retaining nearly the same number of messages per 16-node sub-communicator, leads to an order of magnitude drop in performance.

In overall, the conducted tests demonstrate satisfactory performance and the utility of  the one-sided communication model in Aurora when dealing with highly irregular communication patterns like the one encountered in the FMM application, which has been a comprehensive test platform for one-sided communications in Aurora. The use of one-sided communications in the FMM code opened the path to unprecedented problem sizes due to dedicated efforts of Intel, HPE, and Argonne teams.
\section{Conclusion}\label{Conclusion}
In 2025 Aurora began production operations running scientific applications at exascale using the largest deployment to date of the HPE Slingshot Fabric and Intel's first discrete data center GPU.  Significant work has gone into the deployment and verification of the system and the Slingshot fabric with the resulting machine capabilities demonstrated through MPI benchmark results and scalable benchmarks results including HPL, HPL-MxP, Graph500, and HPCG. Aurora currently ranks as third in the Top500 list at over an exaflop in performance and is the number one ranking machine in the world on the HPL-MxP benchmark. In addition several applications have been demonstrated to scale successfully to a large fraction of the machine including LAMMPS at over 9000 nodes, HACC and AMR-Wind demonstrated to run at 8,000-node scale and Nekbone at 4096 nodes. 
\section{Acknowledgments}\label{Acknowledgment}

This research used resources of the Argonne Leadership Computing Facility, a U.S. Department of Energy (DOE) Office of Science user facility at Argonne National Laboratory and is based on research supported by the U.S. DOE Office of Science-Advanced Scientific Computing Research Program, under Contract No. DE-AC02-06CH11357.

\bibliographystyle{IEEEtran}
\bibliography{references}

\end{document}